
\catcode`\@=11


\message{Loading jyTeX fonts...}



\font\vptrm=cmr5 \font\vptmit=cmmi5 \font\vptsy=cmsy5 \font\vptbf=cmbx5

\skewchar\vptmit='177 \skewchar\vptsy='60 \fontdimen16
\vptsy=\the\fontdimen17 \vptsy

\def\vpt{\ifmmode\err@badsizechange\else
     \@mathfontinit
     \textfont0=\vptrm  \scriptfont0=\vptrm  \scriptscriptfont0=\vptrm
     \textfont1=\vptmit \scriptfont1=\vptmit \scriptscriptfont1=\vptmit
     \textfont2=\vptsy  \scriptfont2=\vptsy  \scriptscriptfont2=\vptsy
     \textfont3=\xptex  \scriptfont3=\xptex  \scriptscriptfont3=\xptex
     \textfont\bffam=\vptbf
     \scriptfont\bffam=\vptbf
     \scriptscriptfont\bffam=\vptbf
     \@fontstyleinit
     \def\rm{\vptrm\fam=\z@}%
     \def\bf{\vptbf\fam=\bffam}%
     \def\oldstyle{\vptmit\fam=\@ne}%
     \rm\fi}


\font\viptrm=cmr6 \font\viptmit=cmmi6 \font\viptsy=cmsy6
\font\viptbf=cmbx6

\skewchar\viptmit='177 \skewchar\viptsy='60 \fontdimen16
\viptsy=\the\fontdimen17 \viptsy

\def\vipt{\ifmmode\err@badsizechange\else
     \@mathfontinit
     \textfont0=\viptrm  \scriptfont0=\vptrm  \scriptscriptfont0=\vptrm
     \textfont1=\viptmit \scriptfont1=\vptmit \scriptscriptfont1=\vptmit
     \textfont2=\viptsy  \scriptfont2=\vptsy  \scriptscriptfont2=\vptsy
     \textfont3=\xptex   \scriptfont3=\xptex  \scriptscriptfont3=\xptex
     \textfont\bffam=\viptbf
     \scriptfont\bffam=\vptbf
     \scriptscriptfont\bffam=\vptbf
     \@fontstyleinit
     \def\rm{\viptrm\fam=\z@}%
     \def\bf{\viptbf\fam=\bffam}%
     \def\oldstyle{\viptmit\fam=\@ne}%
     \rm\fi}

\font\viiptrm=cmr7 \font\viiptmit=cmmi7 \font\viiptsy=cmsy7
\font\viiptit=cmti7 \font\viiptbf=cmbx7

\skewchar\viiptmit='177 \skewchar\viiptsy='60 \fontdimen16
\viiptsy=\the\fontdimen17 \viiptsy

\def\viipt{\ifmmode\err@badsizechange\else
     \@mathfontinit
     \textfont0=\viiptrm  \scriptfont0=\vptrm  \scriptscriptfont0=\vptrm
     \textfont1=\viiptmit \scriptfont1=\vptmit \scriptscriptfont1=\vptmit
     \textfont2=\viiptsy  \scriptfont2=\vptsy  \scriptscriptfont2=\vptsy
     \textfont3=\xptex    \scriptfont3=\xptex  \scriptscriptfont3=\xptex
     \textfont\itfam=\viiptit
     \scriptfont\itfam=\viiptit
     \scriptscriptfont\itfam=\viiptit
     \textfont\bffam=\viiptbf
     \scriptfont\bffam=\vptbf
     \scriptscriptfont\bffam=\vptbf
     \@fontstyleinit
     \def\rm{\viiptrm\fam=\z@}%
     \def\it{\viiptit\fam=\itfam}%
     \def\bf{\viiptbf\fam=\bffam}%
     \def\oldstyle{\viiptmit\fam=\@ne}%
     \rm\fi}


\font\viiiptrm=cmr8 \font\viiiptmit=cmmi8 \font\viiiptsy=cmsy8
\font\viiiptit=cmti8
\font\viiiptbf=cmbx8

\skewchar\viiiptmit='177 \skewchar\viiiptsy='60 \fontdimen16
\viiiptsy=\the\fontdimen17 \viiiptsy

\def\viiipt{\ifmmode\err@badsizechange\else
     \@mathfontinit
     \textfont0=\viiiptrm  \scriptfont0=\viptrm  \scriptscriptfont0=\vptrm
     \textfont1=\viiiptmit \scriptfont1=\viptmit \scriptscriptfont1=\vptmit
     \textfont2=\viiiptsy  \scriptfont2=\viptsy  \scriptscriptfont2=\vptsy
     \textfont3=\xptex     \scriptfont3=\xptex   \scriptscriptfont3=\xptex
     \textfont\itfam=\viiiptit
     \scriptfont\itfam=\viiptit
     \scriptscriptfont\itfam=\viiptit
     \textfont\bffam=\viiiptbf
     \scriptfont\bffam=\viptbf
     \scriptscriptfont\bffam=\vptbf
     \@fontstyleinit
     \def\rm{\viiiptrm\fam=\z@}%
     \def\it{\viiiptit\fam=\itfam}%
     \def\bf{\viiiptbf\fam=\bffam}%
     \def\oldstyle{\viiiptmit\fam=\@ne}%
     \rm\fi}


\def\getixpt{%
     \font\ixptrm=cmr9
     \font\ixptmit=cmmi9
     \font\ixptsy=cmsy9
     \font\ixptit=cmti9
     \font\ixptbf=cmbx9
     \skewchar\ixptmit='177 \skewchar\ixptsy='60
     \fontdimen16 \ixptsy=\the\fontdimen17 \ixptsy}

\def\ixpt{\ifmmode\err@badsizechange\else
     \@mathfontinit
     \textfont0=\ixptrm  \scriptfont0=\viiptrm  \scriptscriptfont0=\vptrm
     \textfont1=\ixptmit \scriptfont1=\viiptmit \scriptscriptfont1=\vptmit
     \textfont2=\ixptsy  \scriptfont2=\viiptsy  \scriptscriptfont2=\vptsy
     \textfont3=\xptex   \scriptfont3=\xptex    \scriptscriptfont3=\xptex
     \textfont\itfam=\ixptit
     \scriptfont\itfam=\viiptit
     \scriptscriptfont\itfam=\viiptit
     \textfont\bffam=\ixptbf
     \scriptfont\bffam=\viiptbf
     \scriptscriptfont\bffam=\vptbf
     \@fontstyleinit
     \def\rm{\ixptrm\fam=\z@}%
     \def\it{\ixptit\fam=\itfam}%
     \def\bf{\ixptbf\fam=\bffam}%
     \def\oldstyle{\ixptmit\fam=\@ne}%
     \rm\fi}


\font\xptrm=cmr10 \font\xptmit=cmmi10 \font\xptsy=cmsy10
\font\xptex=cmex10 \font\xptit=cmti10 \font\xptsl=cmsl10
\font\xptbf=cmbx10 \font\xpttt=cmtt10 \font\xptss=cmss10
\font\xptsc=cmcsc10 \font\xptbfs=cmb10 \font\xptbmit=cmmib10

\skewchar\xptmit='177 \skewchar\xptbmit='177 \skewchar\xptsy='60
\fontdimen16 \xptsy=\the\fontdimen17 \xptsy

\def\xpt{\ifmmode\err@badsizechange\else
     \@mathfontinit
     \textfont0=\xptrm  \scriptfont0=\viiptrm  \scriptscriptfont0=\vptrm
     \textfont1=\xptmit \scriptfont1=\viiptmit \scriptscriptfont1=\vptmit
     \textfont2=\xptsy  \scriptfont2=\viiptsy  \scriptscriptfont2=\vptsy
     \textfont3=\xptex  \scriptfont3=\xptex    \scriptscriptfont3=\xptex
     \textfont\itfam=\xptit
     \scriptfont\itfam=\viiptit
     \scriptscriptfont\itfam=\viiptit
     \textfont\bffam=\xptbf
     \scriptfont\bffam=\viiptbf
     \scriptscriptfont\bffam=\vptbf
     \textfont\bfsfam=\xptbfs
     \scriptfont\bfsfam=\viiptbf
     \scriptscriptfont\bfsfam=\vptbf
     \textfont\bmitfam=\xptbmit
     \scriptfont\bmitfam=\viiptmit
     \scriptscriptfont\bmitfam=\vptmit
     \@fontstyleinit
     \def\rm{\xptrm\fam=\z@}%
     \def\it{\xptit\fam=\itfam}%
     \def\sl{\xptsl}%
     \def\bf{\xptbf\fam=\bffam}%
     \def\tt{\xpttt}%
     \def\ss{\xptss}%
     \def\sc{\xptsc}%
     \def\bfs{\xptbfs\fam=\bfsfam}%
     \def\bmit{\fam=\bmitfam}%
     \def\oldstyle{\xptmit\fam=\@ne}%
     \rm\fi}


\def\getxipt{%
     \font\xiptrm=cmr10  scaled\magstephalf
     \font\xiptmit=cmmi10 scaled\magstephalf
     \font\xiptsy=cmsy10 scaled\magstephalf
     \font\xiptex=cmex10 scaled\magstephalf
     \font\xiptit=cmti10 scaled\magstephalf
     \font\xiptsl=cmsl10 scaled\magstephalf
     \font\xiptbf=cmbx10 scaled\magstephalf
     \font\xipttt=cmtt10 scaled\magstephalf
     \font\xiptss=cmss10 scaled\magstephalf
     \skewchar\xiptmit='177 \skewchar\xiptsy='60
     \fontdimen16 \xiptsy=\the\fontdimen17 \xiptsy}

\def\xipt{\ifmmode\err@badsizechange\else
     \@mathfontinit
     \textfont0=\xiptrm  \scriptfont0=\viiiptrm  \scriptscriptfont0=\viptrm
     \textfont1=\xiptmit \scriptfont1=\viiiptmit \scriptscriptfont1=\viptmit
     \textfont2=\xiptsy  \scriptfont2=\viiiptsy  \scriptscriptfont2=\viptsy
     \textfont3=\xiptex  \scriptfont3=\xptex     \scriptscriptfont3=\xptex
     \textfont\itfam=\xiptit
     \scriptfont\itfam=\viiiptit
     \scriptscriptfont\itfam=\viiptit
     \textfont\bffam=\xiptbf
     \scriptfont\bffam=\viiiptbf
     \scriptscriptfont\bffam=\viptbf
     \@fontstyleinit
     \def\rm{\xiptrm\fam=\z@}%
     \def\it{\xiptit\fam=\itfam}%
     \def\sl{\xiptsl}%
     \def\bf{\xiptbf\fam=\bffam}%
     \def\tt{\xipttt}%
     \def\ss{\xiptss}%
     \def\oldstyle{\xiptmit\fam=\@ne}%
     \rm\fi}


\font\xiiptrm=cmr12 \font\xiiptmit=cmmi12 \font\xiiptsy=cmsy10
scaled\magstep1 \font\xiiptex=cmex10  scaled\magstep1
\font\xiiptit=cmti12 \font\xiiptsl=cmsl12 \font\xiiptbf=cmbx12
\font\xiiptss=cmss12 \font\xiiptsc=cmcsc10 scaled\magstep1
\font\xiiptbfs=cmb10  scaled\magstep1 \font\xiiptbmit=cmmib10
scaled\magstep1

\skewchar\xiiptmit='177 \skewchar\xiiptbmit='177 \skewchar\xiiptsy='60
\fontdimen16 \xiiptsy=\the\fontdimen17 \xiiptsy

\def\xiipt{\ifmmode\err@badsizechange\else
     \@mathfontinit
     \textfont0=\xiiptrm  \scriptfont0=\viiiptrm  \scriptscriptfont0=\viptrm
     \textfont1=\xiiptmit \scriptfont1=\viiiptmit \scriptscriptfont1=\viptmit
     \textfont2=\xiiptsy  \scriptfont2=\viiiptsy  \scriptscriptfont2=\viptsy
     \textfont3=\xiiptex  \scriptfont3=\xptex     \scriptscriptfont3=\xptex
     \textfont\itfam=\xiiptit
     \scriptfont\itfam=\viiiptit
     \scriptscriptfont\itfam=\viiptit
     \textfont\bffam=\xiiptbf
     \scriptfont\bffam=\viiiptbf
     \scriptscriptfont\bffam=\viptbf
     \textfont\bfsfam=\xiiptbfs
     \scriptfont\bfsfam=\viiiptbf
     \scriptscriptfont\bfsfam=\viptbf
     \textfont\bmitfam=\xiiptbmit
     \scriptfont\bmitfam=\viiiptmit
     \scriptscriptfont\bmitfam=\viptmit
     \@fontstyleinit
     \def\rm{\xiiptrm\fam=\z@}%
     \def\it{\xiiptit\fam=\itfam}%
     \def\sl{\xiiptsl}%
     \def\bf{\xiiptbf\fam=\bffam}%
     \def\tt{\xiipttt}%
     \def\ss{\xiiptss}%
     \def\sc{\xiiptsc}%
     \def\bfs{\xiiptbfs\fam=\bfsfam}%
     \def\bmit{\fam=\bmitfam}%
     \def\oldstyle{\xiiptmit\fam=\@ne}%
     \rm\fi}


\def\getxiiipt{%
     \font\xiiiptrm=cmr12  scaled\magstephalf
     \font\xiiiptmit=cmmi12 scaled\magstephalf
     \font\xiiiptsy=cmsy9  scaled\magstep2
     \font\xiiiptit=cmti12 scaled\magstephalf
     \font\xiiiptsl=cmsl12 scaled\magstephalf
     \font\xiiiptbf=cmbx12 scaled\magstephalf
     \font\xiiipttt=cmtt12 scaled\magstephalf
     \font\xiiiptss=cmss12 scaled\magstephalf
     \skewchar\xiiiptmit='177 \skewchar\xiiiptsy='60
     \fontdimen16 \xiiiptsy=\the\fontdimen17 \xiiiptsy}

\def\xiiipt{\ifmmode\err@badsizechange\else
     \@mathfontinit
     \textfont0=\xiiiptrm  \scriptfont0=\xptrm  \scriptscriptfont0=\viiptrm
     \textfont1=\xiiiptmit \scriptfont1=\xptmit \scriptscriptfont1=\viiptmit
     \textfont2=\xiiiptsy  \scriptfont2=\xptsy  \scriptscriptfont2=\viiptsy
     \textfont3=\xivptex   \scriptfont3=\xptex  \scriptscriptfont3=\xptex
     \textfont\itfam=\xiiiptit
     \scriptfont\itfam=\xptit
     \scriptscriptfont\itfam=\viiptit
     \textfont\bffam=\xiiiptbf
     \scriptfont\bffam=\xptbf
     \scriptscriptfont\bffam=\viiptbf
     \@fontstyleinit
     \def\rm{\xiiiptrm\fam=\z@}%
     \def\it{\xiiiptit\fam=\itfam}%
     \def\sl{\xiiiptsl}%
     \def\bf{\xiiiptbf\fam=\bffam}%
     \def\tt{\xiiipttt}%
     \def\ss{\xiiiptss}%
     \def\oldstyle{\xiiiptmit\fam=\@ne}%
     \rm\fi}


\font\xivptrm=cmr12   scaled\magstep1 \font\xivptmit=cmmi12
scaled\magstep1 \font\xivptsy=cmsy10  scaled\magstep2
\font\xivptex=cmex10  scaled\magstep2 \font\xivptit=cmti12
scaled\magstep1 \font\xivptsl=cmsl12  scaled\magstep1
\font\xivptbf=cmbx12  scaled\magstep1
\font\xivptss=cmss12  scaled\magstep1 \font\xivptsc=cmcsc10
scaled\magstep2 \font\xivptbfs=cmb10  scaled\magstep2
\font\xivptbmit=cmmib10 scaled\magstep2

\skewchar\xivptmit='177 \skewchar\xivptbmit='177 \skewchar\xivptsy='60
\fontdimen16 \xivptsy=\the\fontdimen17 \xivptsy

\def\xivpt{\ifmmode\err@badsizechange\else
     \@mathfontinit
     \textfont0=\xivptrm  \scriptfont0=\xptrm  \scriptscriptfont0=\viiptrm
     \textfont1=\xivptmit \scriptfont1=\xptmit \scriptscriptfont1=\viiptmit
     \textfont2=\xivptsy  \scriptfont2=\xptsy  \scriptscriptfont2=\viiptsy
     \textfont3=\xivptex  \scriptfont3=\xptex  \scriptscriptfont3=\xptex
     \textfont\itfam=\xivptit
     \scriptfont\itfam=\xptit
     \scriptscriptfont\itfam=\viiptit
     \textfont\bffam=\xivptbf
     \scriptfont\bffam=\xptbf
     \scriptscriptfont\bffam=\viiptbf
     \textfont\bfsfam=\xivptbfs
     \scriptfont\bfsfam=\xptbfs
     \scriptscriptfont\bfsfam=\viiptbf
     \textfont\bmitfam=\xivptbmit
     \scriptfont\bmitfam=\xptbmit
     \scriptscriptfont\bmitfam=\viiptmit
     \@fontstyleinit
     \def\rm{\xivptrm\fam=\z@}%
     \def\it{\xivptit\fam=\itfam}%
     \def\sl{\xivptsl}%
     \def\bf{\xivptbf\fam=\bffam}%
     \def\tt{\xivpttt}%
     \def\ss{\xivptss}%
     \def\sc{\xivptsc}%
     \def\bfs{\xivptbfs\fam=\bfsfam}%
     \def\bmit{\fam=\bmitfam}%
     \def\oldstyle{\xivptmit\fam=\@ne}%
     \rm\fi}


\font\xviiptrm=cmr17 \font\xviiptmit=cmmi12 scaled\magstep2
\font\xviiptsy=cmsy10 scaled\magstep3 \font\xviiptex=cmex10
scaled\magstep3 \font\xviiptit=cmti12 scaled\magstep2
\font\xviiptbf=cmbx12 scaled\magstep2 \font\xviiptbfs=cmb10
scaled\magstep3

\skewchar\xviiptmit='177 \skewchar\xviiptsy='60 \fontdimen16
\xviiptsy=\the\fontdimen17 \xviiptsy

\def\xviipt{\ifmmode\err@badsizechange\else
     \@mathfontinit
     \textfont0=\xviiptrm  \scriptfont0=\xiiptrm  \scriptscriptfont0=\viiiptrm
     \textfont1=\xviiptmit \scriptfont1=\xiiptmit \scriptscriptfont1=\viiiptmit
     \textfont2=\xviiptsy  \scriptfont2=\xiiptsy  \scriptscriptfont2=\viiiptsy
     \textfont3=\xviiptex  \scriptfont3=\xiiptex  \scriptscriptfont3=\xptex
     \textfont\itfam=\xviiptit
     \scriptfont\itfam=\xiiptit
     \scriptscriptfont\itfam=\viiiptit
     \textfont\bffam=\xviiptbf
     \scriptfont\bffam=\xiiptbf
     \scriptscriptfont\bffam=\viiiptbf
     \textfont\bfsfam=\xviiptbfs
     \scriptfont\bfsfam=\xiiptbfs
     \scriptscriptfont\bfsfam=\viiiptbf
     \@fontstyleinit
     \def\rm{\xviiptrm\fam=\z@}%
     \def\it{\xviiptit\fam=\itfam}%
     \def\bf{\xviiptbf\fam=\bffam}%
     \def\bfs{\xviiptbfs\fam=\bfsfam}%
     \def\oldstyle{\xviiptmit\fam=\@ne}%
     \rm\fi}


\font\xxiptrm=cmr17  scaled\magstep1


\def\xxipt{\ifmmode\err@badsizechange\else
     \@mathfontinit
     \@fontstyleinit
     \def\rm{\xxiptrm\fam=\z@}%
     \rm\fi}


\font\xxvptrm=cmr17  scaled\magstep2


\def\xxvpt{\ifmmode\err@badsizechange\else
     \@mathfontinit
     \@fontstyleinit
     \def\rm{\xxvptrm\fam=\z@}%
     \rm\fi}




\message{Loading jyTeX macros...}

\message{modifications to plain.tex,}


\def\newcount{\alloc@0\count\countdef\insc@unt}
\def\newdimen{\alloc@1\dimen\dimendef\insc@unt}
\def\newskip{\alloc@2\skip\skipdef\insc@unt}
\def\newmuskip{\alloc@3\muskip\muskipdef\@cclvi}
\def\newbox{\alloc@4\box\chardef\insc@unt}
\def\newtoks{\alloc@5\toks\toksdef\@cclvi}
\def\newhelp#1#2{\newtoks#1\global#1\expandafter{\csname#2\endcsname}}
\def\newread{\alloc@6\read\chardef\sixt@@n}
\def\newwrite{\alloc@7\write\chardef\sixt@@n}
\def\newfam{\alloc@8\fam\chardef\sixt@@n}
\def\newinsert#1{\global\advance\insc@unt by\m@ne
     \ch@ck0\insc@unt\count
     \ch@ck1\insc@unt\dimen
     \ch@ck2\insc@unt\skip
     \ch@ck4\insc@unt\box
     \allocationnumber=\insc@unt
     \global\chardef#1=\allocationnumber
     \wlog{\string#1=\string\insert\the\allocationnumber}}
\def\newif#1{\count@\escapechar \escapechar\m@ne
     \expandafter\expandafter\expandafter
          \xdef\@if#1{true}{\let\noexpand#1=\noexpand\iftrue}%
     \expandafter\expandafter\expandafter
          \xdef\@if#1{false}{\let\noexpand#1=\noexpand\iffalse}%
     \global\@if#1{false}\escapechar=\count@}


\newlinechar=`\^^J
\overfullrule=0pt




\let\itfam=\undefined

\let\bffam=\undefined

\count18=3


\chardef\sharps="19


\mathchardef\alpha="710B \mathchardef\beta="710C \mathchardef\gamma="710D
\mathchardef\delta="710E \mathchardef\epsilon="710F
\mathchardef\zeta="7110 \mathchardef\eta="7111 \mathchardef\theta="7112
\mathchardef\iota="7113 \mathchardef\kappa="7114
\mathchardef\lambda="7115 \mathchardef\mu="7116 \mathchardef\nu="7117
\mathchardef\xi="7118 \mathchardef\pi="7119 \mathchardef\rho="711A
\mathchardef\sigma="711B \mathchardef\tau="711C
\mathchardef\upsilon="711D \mathchardef\phi="711E \mathchardef\chi="711F
\mathchardef\psi="7120 \mathchardef\omega="7121
\mathchardef\varepsilon="7122 \mathchardef\vartheta="7123
\mathchardef\varpi="7124 \mathchardef\varrho="7125
\mathchardef\varsigma="7126 \mathchardef\varphi="7127
\mathchardef\imath="717B \mathchardef\jmath="717C \mathchardef\ell="7160
\mathchardef\wp="717D \mathchardef\partial="7140 \mathchardef\flat="715B
\mathchardef\natural="715C \mathchardef\sharp="715D



\def\angle{{\vbox{\ialign{$\m@th\scriptstyle##$\crcr
     \not\mathrel{\mkern14mu}\crcr
     \noalign{\nointerlineskip}
     \mkern2.5mu\leaders\hrule height.34\rp@\hfill\mkern2.5mu\crcr}}}}
\def\vdots{\vbox{\baselineskip4\rp@ \lineskiplimit\z@
     \kern6\rp@\hbox{.}\hbox{.}\hbox{.}}}
\def\ddots{\mathinner{\mkern1mu\raise7\rp@\vbox{\kern7\rp@\hbox{.}}\mkern2mu
     \raise4\rp@\hbox{.}\mkern2mu\raise\rp@\hbox{.}\mkern1mu}}
\def\overrightarrow#1{\vbox{\ialign{##\crcr
     \rightarrowfill\crcr
     \noalign{\kern-\rp@\nointerlineskip}
     $\hfil\displaystyle{#1}\hfil$\crcr}}}
\def\overleftarrow#1{\vbox{\ialign{##\crcr
     \leftarrowfill\crcr
     \noalign{\kern-\rp@\nointerlineskip}
     $\hfil\displaystyle{#1}\hfil$\crcr}}}
\def\overbrace#1{\mathop{\vbox{\ialign{##\crcr
     \noalign{\kern3\rp@}
     \downbracefill\crcr
     \noalign{\kern3\rp@\nointerlineskip}
     $\hfil\displaystyle{#1}\hfil$\crcr}}}\limits}
\def\underbrace#1{\mathop{\vtop{\ialign{##\crcr
     $\hfil\displaystyle{#1}\hfil$\crcr
     \noalign{\kern3\rp@\nointerlineskip}
     \upbracefill\crcr
     \noalign{\kern3\rp@}}}}\limits}
\def\big#1{{\hbox{$\left#1\vbox to8.5\rp@ {}\right.\n@space$}}}
\def\Big#1{{\hbox{$\left#1\vbox to11.5\rp@ {}\right.\n@space$}}}
\def\bigg#1{{\hbox{$\left#1\vbox to14.5\rp@ {}\right.\n@space$}}}
\def\Bigg#1{{\hbox{$\left#1\vbox to17.5\rp@ {}\right.\n@space$}}}
\def\@vereq#1#2{\lower.5\rp@\vbox{\baselineskip\z@skip\lineskip-.5\rp@
     \ialign{$\m@th#1\hfil##\hfil$\crcr#2\crcr=\crcr}}}
\def\rlh@#1{\vcenter{\hbox{\ooalign{\raise2\rp@
     \hbox{$#1\rightharpoonup$}\crcr
     $#1\leftharpoondown$}}}}
\def\bordermatrix#1{\begingroup\m@th
     \setbox\z@\vbox{%
          \def\cr{\crcr\noalign{\kern2\rp@\global\let\cr\endline}}%
          \ialign{$##$\hfil\kern2\rp@\kern\p@renwd
               &\thinspace\hfil$##$\hfil&&\quad\hfil$##$\hfil\crcr
               \omit\strut\hfil\crcr
               \noalign{\kern-\baselineskip}%
               #1\crcr\omit\strut\cr}}%
     \setbox\tw@\vbox{\unvcopy\z@\global\setbox\@ne\lastbox}%
     \setbox\tw@\hbox{\unhbox\@ne\unskip\global\setbox\@ne\lastbox}%
     \setbox\tw@\hbox{$\kern\wd\@ne\kern-\p@renwd\left(\kern-\wd\@ne
          \global\setbox\@ne\vbox{\box\@ne\kern2\rp@}%
          \vcenter{\kern-\ht\@ne\unvbox\z@\kern-\baselineskip}%
          \,\right)$}%
     \null\;\vbox{\kern\ht\@ne\box\tw@}\endgroup}
\def\endinsert{\egroup
     \if@mid\dimen@\ht\z@
          \advance\dimen@\dp\z@
          \advance\dimen@12\rp@
          \advance\dimen@\pagetotal
          \ifdim\dimen@>\pagegoal\@midfalse\p@gefalse\fi
     \fi
     \if@mid\bigskip\box\z@
          \bigbreak
     \else\insert\topins{\penalty100 \splittopskip\z@skip
               \splitmaxdepth\maxdimen\floatingpenalty\z@
               \ifp@ge\dimen@\dp\z@
                    \vbox to\vsize{\unvbox\z@\kern-\dimen@}%
               \else\box\z@\nobreak\bigskip
               \fi}%
     \fi
     \endgroup}


\def\cases#1{\left\{\,\vcenter{\m@th
     \ialign{$##\hfil$&\quad##\hfil\crcr#1\crcr}}\right.}
\def\matrix#1{\null\,\vcenter{\m@th
     \ialign{\hfil$##$\hfil&&\quad\hfil$##$\hfil\crcr
          \mathstrut\crcr
          \noalign{\kern-\baselineskip}
          #1\crcr
          \mathstrut\crcr
          \noalign{\kern-\baselineskip}}}\,}


\newif\ifraggedbottom

\def\raggedbottom{\ifraggedbottom\else
     \advance\topskip by\z@ plus60pt \raggedbottomtrue\fi}%
\def\normalbottom{\ifraggedbottom
     \advance\topskip by\z@ plus-60pt \raggedbottomfalse\fi}

\message{hacks,}


\toksdef\toks@i=1 \toksdef\toks@ii=2


\def\TeX{T\kern-.1667em \lower.5ex \hbox{E}\kern-.125em X\null}
\def\jyTeX{{\leavevmode
     \raise.587ex \hbox{\it\j}\kern-.1em \lower.048ex \hbox{\it y}\kern-.12em
     \TeX}}

\let\then=\iftrue
\def\ifnoarg#1\then{\def\hack@{#1}\ifx\hack@\empty}
\def\ifundefined#1\then{%
     \expandafter\ifx\csname\expandafter\blank\string#1\endcsname\relax}
\def\useif#1\then{\csname#1\endcsname}
\def\usename#1{\csname#1\endcsname}
\def\useafter#1#2{\expandafter#1\csname#2\endcsname}

\long\def\loop#1\repeat{\def\@iterate{#1\expandafter\@iterate\fi}\@iterate
     \let\@iterate=\relax}

\let\TeXend=\end
\def\begin#1{\begingroup\def\@@blockname{#1}\usename{begin#1}}
\def\end#1{\usename{end#1}\def\hack@{#1}%
     \ifx\@@blockname\hack@
          \endgroup
     \else\err@badgroup\hack@\@@blockname
     \fi}
\def\@@blockname{}

\def\defaultoption[#1]#2{%
     \def\hack@{\ifx\hack@ii[\toks@={#2}\else\toks@={#2[#1]}\fi\the\toks@}%
     \futurelet\hack@ii\hack@}

\def\markup#1{\let\@@marksf=\empty
     \ifhmode\edef\@@marksf{\spacefactor=\the\spacefactor\relax}\/\fi
     ${}^{\hbox{\subscriptfonts#1}}$\@@marksf}


\newtoks\shortyear
\newtoks\militaryhour
\newtoks\standardhour
\newtoks\minute
\newtoks\amorpm

\def\settime{\count@=\time\divide\count@ by60
     \militaryhour=\expandafter{\number\count@}%
     {\multiply\count@ by-60 \advance\count@ by\time
          \xdef\hack@{\ifnum\count@<10 0\fi\number\count@}}%
     \minute=\expandafter{\hack@}%
     \ifnum\count@<12
          \amorpm={am}
     \else\amorpm={pm}
          \ifnum\count@>12 \advance\count@ by-12 \fi
     \fi
     \standardhour=\expandafter{\number\count@}%
     \def\hack@19##1##2{\shortyear={##1##2}}%
          \expandafter\hack@\the\year}

\def\monthword#1{%
     \ifcase#1
          $\bullet$\err@badcountervalue{monthword}%
          \or January\or February\or March\or April\or May\or June%
          \or July\or August\or September\or October\or November\or December%
     \else$\bullet$\err@badcountervalue{monthword}%
     \fi}

\def\monthabbr#1{%
     \ifcase#1
          $\bullet$\err@badcountervalue{monthabbr}%
          \or Jan\or Feb\or Mar\or Apr\or May\or Jun%
          \or Jul\or Aug\or Sep\or Oct\or Nov\or Dec%
     \else$\bullet$\err@badcountervalue{monthabbr}%
     \fi}

\def\militarytime{\the\militaryhour:\the\minute}
\def\standardtime{\the\standardhour:\the\minute}


\def\@setnumstyle#1#2{\expandafter\global\expandafter\expandafter
     \expandafter\let\expandafter\expandafter
     \csname @\expandafter\blank\string#1style\endcsname
     \csname#2\endcsname}
\def\numstyle#1{\usename{@\expandafter\blank\string#1style}#1}
\def\ifblank#1\then{\useafter\ifx{@\expandafter\blank\string#1}\blank}

\def\blank#1{}

\def\Roman#1{\expandafter\uppercase\expandafter{\romannumeral#1}}
\def\alphabetic#1{%
     \ifcase#1
          $\bullet$\err@badcountervalue{alphabetic}%
          \or a\or b\or c\or d\or e\or f\or g\or h\or i\or j\or k\or l\or m%
          \or n\or o\or p\or q\or r\or s\or t\or u\or v\or w\or x\or y\or z%
     \else$\bullet$\err@badcountervalue{alphabetic}%
     \fi}
\def\Alphabetic#1{\expandafter\uppercase\expandafter{\alphabetic{#1}}}
\def\symbols#1{%
     \ifcase#1
          $\bullet$\err@badcountervalue{symbols}%
          \or*\or\dag\or\ddag\or\S\or$\|$%
          \or**\or\dag\dag\or\ddag\ddag\or\S\S\or$\|\|$%
     \else$\bullet$\err@badcountervalue{symbols}%
     \fi}


\catcode`\^^?=13 \def^^?{\relax}

\def\trimleading#1\to#2{\edef#2{#1}%
     \expandafter\@trimleading\expandafter#2#2^^?^^?}
\def\@trimleading#1#2#3^^?{\ifx#2^^?\def#1{}\else\def#1{#2#3}\fi}

\def\trimtrailing#1\to#2{\edef#2{#1}%
     \expandafter\@trimtrailing\expandafter#2#2^^? ^^?\relax}
\def\@trimtrailing#1#2 ^^?#3{\ifx#3\relax\toks@={}%
     \else\def#1{#2}\toks@={\trimtrailing#1\to#1}\fi
     \the\toks@}

\def\trim#1\to#2{\trimleading#1\to#2\trimtrailing#2\to#2}

\catcode`\^^?=15


\long\def\additemL#1\to#2{\toks@={\^^\{#1}}\toks@ii=\expandafter{#2}%
     \xdef#2{\the\toks@\the\toks@ii}}

\long\def\additemR#1\to#2{\toks@={\^^\{#1}}\toks@ii=\expandafter{#2}%
     \xdef#2{\the\toks@ii\the\toks@}}

\def\getitemL#1\to#2{\expandafter\@getitemL#1\hack@#1#2}
\def\@getitemL\^^\#1#2\hack@#3#4{\def#4{#1}\def#3{#2}}

\message{font macros,}


\newdimen\rp@
\newcount\@@sizeindex \@@sizeindex=0
\newcount\@@factori
\newcount\@@factorii
\newcount\@@factoriii
\newcount\@@factoriv

\countdef\maxfam=18
\newfam\itfam
\newfam\bffam
\newfam\bfsfam
\newfam\bmitfam

\def\@mathfontinit{\count@=4
     \loop\textfont\count@=\nullfont
          \scriptfont\count@=\nullfont
          \scriptscriptfont\count@=\nullfont
          \ifnum\count@<\maxfam\advance\count@ by\@ne
     \repeat}

\def\@fontstyleinit{%
     \def\it{\err@fontnotavailable\it}%
     \def\bf{\err@fontnotavailable\bf}%
     \def\bfs{\err@bfstobf}%
     \def\bmit{\err@fontnotavailable\bmit}%
     \def\sc{\err@fontnotavailable\sc}%
     \def\sl{\err@sltoit}%
     \def\ss{\err@fontnotavailable\ss}%
     \def\tt{\err@fontnotavailable\tt}}

\def\@parameterinit#1{\rm\rp@=.1em \@getscaling{#1}%
     \let\^^\=\@doscaling\scalingskipslist
     \setbox\strutbox=\hbox{\vrule
          height.708\baselineskip depth.292\baselineskip width\z@}}

\def\@getfactor#1#2#3#4{\@@factori=#1 \@@factorii=#2
     \@@factoriii=#3 \@@factoriv=#4}

\def\@getscaling#1{\count@=#1 \advance\count@ by-\@@sizeindex\@@sizeindex=#1
     \ifnum\count@<0
          \let\@mulordiv=\divide
          \let\@divormul=\multiply
          \multiply\count@ by\m@ne
     \else\let\@mulordiv=\multiply
          \let\@divormul=\divide
     \fi
     \edef\@@scratcha{\ifcase\count@                {1}{1}{1}{1}\or
          {1}{7}{23}{3}\or     {2}{5}{3}{1}\or      {9}{89}{13}{1}\or
          {6}{25}{6}{1}\or     {8}{71}{14}{1}\or    {6}{25}{36}{5}\or
          {1}{7}{53}{4}\or     {12}{125}{108}{5}\or {3}{14}{53}{5}\or
          {6}{41}{17}{1}\or    {13}{31}{13}{2}\or   {9}{107}{71}{2}\or
          {11}{139}{124}{3}\or {1}{6}{43}{2}\or     {10}{107}{42}{1}\or
          {1}{5}{43}{2}\or     {5}{69}{65}{1}\or    {11}{97}{91}{2}\fi}%
     \expandafter\@getfactor\@@scratcha}

\def\@doscaling#1{\@mulordiv#1by\@@factori\@divormul#1by\@@factorii
     \@mulordiv#1by\@@factoriii\@divormul#1by\@@factoriv}


\newskip\headskip
\newskip\footskip

\def\typesize=#1pt{\count@=#1 \advance\count@ by-10
     \ifcase\count@
          \@setsizex\or\err@badtypesize\or
          \@setsizexii\or\err@badtypesize\or
          \@setsizexiv
     \else\err@badtypesize
     \fi}

\def\@setsizex{\getixpt
     \def\subsubscriptfonts{\vpt}%
          \def\subsubscriptsize{\vpt\@parameterinit{-8}}%
     \def\subscriptfonts{\viipt}\def\subscriptsize{\viipt\@parameterinit{-4}}%
     \def\footnotefonts{\viiipt}\def\footnotesize{\viiipt\@parameterinit{-2}}%
     \def\smallfonts{\ixpt}\def\smallsize{\ixpt\@parameterinit{-1}}%
     \def\normalfonts{\xpt}\def\normalsize{\xpt\@parameterinit{0}}%
     \def\bigfonts{\xiipt}\def\bigsize{\xiipt\@parameterinit{2}}%
     \def\Bigfonts{\xivpt}\def\Bigsize{\xivpt\@parameterinit{4}}%
     \def\biggfonts{\xviipt}\def\biggsize{\xviipt\@parameterinit{6}}%
     \def\Biggfonts{\xxipt}\def\Biggsize{\xxipt\@parameterinit{8}}%
     \def\tinyfonts{\vpt}\def\tinysize{\vpt\@parameterinit{-8}}%
     \def\HUGEFONTS{\xxvpt}\def\HUGESIZE{\xxvpt\@parameterinit{10}}%
     \normalsize\fixedskipslist}

\def\@setsizexii{\getxipt
     \def\subsubscriptfonts{\vipt}%
          \def\subsubscriptsize{\vipt\@parameterinit{-6}}%
     \def\subscriptfonts{\viiipt}%
          \def\subscriptsize{\viiipt\@parameterinit{-2}}%
     \def\footnotefonts{\xpt}\def\footnotesize{\xpt\@parameterinit{0}}%
     \def\smallfonts{\xipt}\def\smallsize{\xipt\@parameterinit{1}}%
     \def\normalfonts{\xiipt}\def\normalsize{\xiipt\@parameterinit{2}}%
     \def\bigfonts{\xivpt}\def\bigsize{\xivpt\@parameterinit{4}}%
     \def\Bigfonts{\xviipt}\def\Bigsize{\xviipt\@parameterinit{6}}%
     \def\biggfonts{\xxipt}\def\biggsize{\xxipt\@parameterinit{8}}%
     \def\Biggfonts{\xxvpt}\def\Biggsize{\xxvpt\@parameterinit{10}}%
     \def\tinyfonts{\vpt}\def\tinysize{\vpt\@parameterinit{-8}}%
     \def\HUGEFONTS{\xxvpt}\def\HUGESIZE{\xxvpt\@parameterinit{10}}%
     \normalsize\fixedskipslist}

\def\@setsizexiv{\getxiiipt
     \def\subsubscriptfonts{\viipt}%
          \def\subsubscriptsize{\viipt\@parameterinit{-4}}%
     \def\subscriptfonts{\xpt}\def\subscriptsize{\xpt\@parameterinit{0}}%
     \def\footnotefonts{\xiipt}\def\footnotesize{\xiipt\@parameterinit{2}}%
     \def\smallfonts{\xiiipt}\def\smallsize{\xiiipt\@parameterinit{3}}%
     \def\normalfonts{\xivpt}\def\normalsize{\xivpt\@parameterinit{4}}%
     \def\bigfonts{\xviipt}\def\bigsize{\xviipt\@parameterinit{6}}%
     \def\Bigfonts{\xxipt}\def\Bigsize{\xxipt\@parameterinit{8}}%
     \def\biggfonts{\xxvpt}\def\biggsize{\xxvpt\@parameterinit{10}}%
     \def\Biggfonts{\err@sizetoolarge\Biggfonts\HUGEFONTS}%
          \def\Biggsize{\err@sizetoolarge\Biggsize\HUGESIZE}%
     \def\tinyfonts{\vpt}\def\tinysize{\vpt\@parameterinit{-8}}%
     \def\HUGEFONTS{\xxvpt}\def\HUGESIZE{\xxvpt\@parameterinit{10}}%
     \normalsize\fixedskipslist}

\def\subsubscriptfonts{\vpt} \def\subsubscriptsize{\vpt\@parameterinit{-8}}
\def\subscriptfonts{\viipt}  \def\subscriptsize{\viipt\@parameterinit{-4}}
\def\footnotefonts{\viiipt}  \def\footnotesize{\viiipt\@parameterinit{-2}}
\def\smallfonts{\err@sizenotavailable\smallfonts}
                             \def\smallsize{\ixpt\@parameterinit{-1}}
\def\normalfonts{\xpt}       \def\normalsize{\xpt\@parameterinit{0}}
\def\bigfonts{\xiipt}        \def\bigsize{\xiipt\@parameterinit{2}}
\def\Bigfonts{\xivpt}        \def\Bigsize{\xivpt\@parameterinit{4}}
\def\biggfonts{\xviipt}      \def\biggsize{\xviipt\@parameterinit{6}}
\def\Biggfonts{\xxipt}       \def\Biggsize{\xxipt\@parameterinit{8}}
\def\tinyfonts{\vpt}         \def\tinysize{\vpt\@parameterinit{-8}}
\def\HUGEFONTS{\xxvpt}       \def\HUGESIZE{\xxvpt\@parameterinit{10}}

\message{document layout,}


\newtoks\everyoutput \everyoutput={}
\newdimen\depthofpage
\newcount\pagenum \pagenum=0

\newdimen\oddtopmargin  \newdimen\eventopmargin
\newdimen\oddleftmargin \newdimen\evenleftmargin
\newtoks\oddhead        \newtoks\evenhead
\newtoks\oddfoot        \newtoks\evenfoot

\def\topmargin{\afterassignment\@seteventop\oddtopmargin}
\def\leftmargin{\afterassignment\@setevenleft\oddleftmargin}
\def\head{\afterassignment\@setevenhead\oddhead}
\def\foot{\afterassignment\@setevenfoot\oddfoot}

\def\@seteventop{\eventopmargin=\oddtopmargin}
\def\@setevenleft{\evenleftmargin=\oddleftmargin}
\def\@setevenhead{\evenhead=\oddhead}
\def\@setevenfoot{\evenfoot=\oddfoot}

\def\pagenumstyle#1{\@setnumstyle\pagenum{#1}}

\newif\ifdraft
\def\draft{\drafttrue\leftmargin=.5in \overfullrule=5pt }

\def\outputstyle#1{\global\expandafter\let\expandafter
          \@outputstyle\csname#1output\endcsname
     \usename{#1setup}}

\output={\@outputstyle}

\def\normaloutput{\the\everyoutput
     \global\advance\pagenum by\@ne
     \ifodd\pagenum
          \voffset=\oddtopmargin \hoffset=\oddleftmargin
     \else\voffset=\eventopmargin \hoffset=\evenleftmargin
     \fi
     \advance\voffset by-1in  \advance\hoffset by-1in
     \count0=\pagenum
     \expandafter\shipout\pagebox
     \ifnum\outputpenalty>-\@MM\else\dosupereject\fi}

\newdimen\fullhsize
\newbox\leftpage
\newcount\leftpagenum
\newcount\outputpagenum \outputpagenum=0
\let\leftorright=L

\def\twoupoutput{\the\everyoutput
     \global\advance\pagenum by\@ne
     \if L\leftorright
          \global\setbox\leftpage=\leftline{\pagebox}%
          \global\leftpagenum=\pagenum
          \global\let\leftorright=R%
     \else\global\advance\outputpagenum by\@ne
          \ifodd\outputpagenum
               \voffset=\oddtopmargin \hoffset=\oddleftmargin
          \else\voffset=\eventopmargin \hoffset=\evenleftmargin
          \fi
          \advance\voffset by-1in  \advance\hoffset by-1in
          \count0=\leftpagenum \count1=\pagenum
          \shipout\vbox{\hbox to\fullhsize
               {\box\leftpage\hfil\leftline{\pagebox}}}%
          \global\let\leftorright=L%
     \fi
     \ifnum\outputpenalty>-\@MM
     \else\dosupereject
          \if R\leftorright
               \globaldefs=\@ne\head={\hfil}\foot={\hfil}\globaldefs=\z@
               \null\newpage
          \fi
     \fi}

\def\pagebox{\vbox{\makeheadline\pagebody\makefootline}}

\def\makeheadline{%
     \vbox to\z@{\baselinestretch=\@m
          \vskip\topskip\vskip-.708\baselineskip\vskip-\headskip
          \line{\vbox to\ht\strutbox{}%
               \ifodd\pagenum\the\oddhead\else\the\evenhead\fi}%
          \vss}%
     \nointerlineskip}

\def\pagebody{\vbox to\vsize{%
     \boxmaxdepth\maxdepth
     \ifvoid\topins\else\unvbox\topins\fi
     \depthofpage=\dp255
     \unvbox255
     \ifraggedbottom\kern-\depthofpage\vfil\fi
     \ifvoid\footins
     \else\vskip\skip\footins
          \footnoterule
          \unvbox\footins
          \vskip-\footnoteskip
     \fi}}

\def\makefootline{\baselineskip=\footskip
     \line{\ifodd\pagenum\the\oddfoot\else\the\evenfoot\fi}}


\newskip\abovechapterskip
\newskip\belowchapterskip
\newskip\abovesectionskip
\newskip\belowsectionskip
\newskip\abovesubsectionskip
\newskip\belowsubsectionskip

\def\chapterstyle#1{\global\expandafter\let\expandafter\@chapterstyle
     \csname#1text\endcsname}
\def\sectionstyle#1{\global\expandafter\let\expandafter\@sectionstyle
     \csname#1text\endcsname}
\def\subsectionstyle#1{\global\expandafter\let\expandafter\@subsectionstyle
     \csname#1text\endcsname}

\def\chapter#1{%
     \ifdim\lastskip=17sp \else\chapterbreak\vskip\abovechapterskip\fi
     \@chapterstyle{\ifblank\chapternumstyle\then
          \else\newchapternum=\next\chapternumformat\ \fi#1}%
     \nobreak\vskip\belowchapterskip\vskip17sp }

\def\section#1{%
     \ifdim\lastskip=17sp \else\sectionbreak\vskip\abovesectionskip\fi
     \@sectionstyle{\ifblank\sectionnumstyle\then
          \else\newsectionnum=\next\sectionnumformat\ \fi#1}%
     \nobreak\vskip\belowsectionskip\vskip17sp }

\def\subsection#1{%
     \ifdim\lastskip=17sp \else\subsectionbreak\vskip\abovesubsectionskip\fi
     \@subsectionstyle{\ifblank\subsectionnumstyle\then
          \else\newsubsectionnum=\next\subsectionnumformat\ \fi#1}%
     \nobreak\vskip\belowsubsectionskip\vskip17sp }


\let\TeXunderline=\underline
\let\TeXoverline=\overline
\def\underline#1{\relax\ifmmode\TeXunderline{#1}\else
     $\TeXunderline{\hbox{#1}}$\fi}
\def\overline#1{\relax\ifmmode\TeXoverline{#1}\else
     $\TeXoverline{\hbox{#1}}$\fi}

\def\baselinestretch{\afterassignment\@baselinestretch\count@}
\def\@baselinestretch{\baselineskip=\normalbaselineskip
     \divide\baselineskip by\@m\baselineskip=\count@\baselineskip
     \setbox\strutbox=\hbox{\vrule
          height.708\baselineskip depth.292\baselineskip width\z@}%
     \bigskipamount=\the\baselineskip
          plus.25\baselineskip minus.25\baselineskip
     \medskipamount=.5\baselineskip
          plus.125\baselineskip minus.125\baselineskip
     \smallskipamount=.25\baselineskip
          plus.0625\baselineskip minus.0625\baselineskip}

\def\\{\ifhmode\ifnum\lastpenalty=-\@M\else\hfil\penalty-\@M\fi\fi
     \ignorespaces}
\def\newpage{\vfil\break}

\def\lefttext#1{\par{\@text\leftskip=\z@\rightskip=\centering
     \noindent#1\par}}
\def\righttext#1{\par{\@text\leftskip=\centering\rightskip=\z@
     \noindent#1\par}}
\def\centertext#1{\par{\@text\leftskip=\centering\rightskip=\centering
     \noindent#1\par}}
\def\@text{\parindent=\z@ \parfillskip=\z@ \everypar={}%
     \spaceskip=.3333em \xspaceskip=.5em
     \def\\{\ifhmode\ifnum\lastpenalty=-\@M\else\penalty-\@M\fi\fi
          \ignorespaces}}

\def\beginleft{\par\@text\leftskip=\z@ \rightskip=\centering}
     
\def\beginright{\par\@text\leftskip=\centering\rightskip=\z@ }
     
\def\begincenter{\par\@text\leftskip=\centering\rightskip=\centering}

\def\beginnarrow{\defaultoption[\parindent]\@beginnarrow}
\def\@beginnarrow[#1]{\par\advance\leftskip by#1\advance\rightskip by#1}

\begingroup
\catcode`\[=1 \catcode`\{=11 \gdef\beginignore[\endgroup\bgroup
     \catcode`\e=0 \catcode`\\=12 \catcode`\{=11 \catcode`\f=12 \let\or=\relax
     \let\nd{ignor=\fi \let\}=\egroup
     \iffalse}
\endgroup

\long\def\marginnote#1{\leavevmode
     \edef\@marginsf{\spacefactor=\the\spacefactor\relax}%
     \ifdraft\strut\vadjust{%
          \hbox to\z@{\hskip\hsize\hskip.1in
               \vbox to\z@{\vskip-\dp\strutbox
                    \marginnoteformat
                    \vskip-\ht\strutbox
                    \noindent\strut#1\par
                    \vss}%
               \hss}}%
     \fi
     \@marginsf}


\newtoks\everybye \everybye={\par\vfil}
\outer\def\bye{\the\everybye
     \footnotecheck
     \prelabelcheck
     \streamcheck
     \supereject
     \TeXend}

\message{footnotes,}

\newcount\footnotenum \footnotenum=0
\newskip\footnoteskip
\let\@footnotelist=\empty

\def\footnotenumstyle#1{\@setnumstyle\footnotenum{#1}%
     \useafter\ifx{@footnotenumstyle}\symbols
          \global\let\@footup=\empty
     \else\global\let\@footup=\markup
     \fi}

\def\footnote{\footnotecheck\defaultoption[]\@footnote}
\def\@footnote[#1]{\@footnotemark[#1]\@footnotetext}

\def\footnotemark{\defaultoption[]\@footnotemark}
\def\@footnotemark[#1]{\let\@footsf=\empty
     \ifhmode\edef\@footsf{\spacefactor=\the\spacefactor\relax}\/\fi
     \ifnoarg#1\then
          \global\advance\footnotenum by\@ne
          \@footup{\footnotenumformat}%
          \edef\@@foota{\footnotenum=\the\footnotenum\relax}%
          \expandafter\additemR\expandafter\@footup\expandafter
               {\@@foota\footnotenumformat}\to\@footnotelist
          \global\let\@footnotelist=\@footnotelist
     \else\markup{#1}%
          \additemR\markup{#1}\to\@footnotelist
          \global\let\@footnotelist=\@footnotelist
     \fi
     \@footsf}

\def\footnotetext{%
     \ifx\@footnotelist\empty\err@extrafootnotetext\else\@footnotetext\fi}
\def\@footnotetext{%
     \getitemL\@footnotelist\to\@@foota
     \global\let\@footnotelist=\@footnotelist
     \insert\footins\bgroup
     \footnoteformat
     \splittopskip=\ht\strutbox\splitmaxdepth=\dp\strutbox
     \interlinepenalty=\interfootnotelinepenalty\floatingpenalty=\@MM
     \noindent\llap{\@@foota}\strut
     \bgroup\aftergroup\@footnoteend
     \let\@@scratcha=}
\def\@footnoteend{\strut\par\vskip\footnoteskip\egroup}

\def\footnoterule{\normalfonts
     \kern-.3em \hrule width2in height.04em \kern .26em }

\def\footnotecheck{%
     \ifx\@footnotelist\empty
     \else\err@extrafootnotemark
          \global\let\@footnotelist=\empty
     \fi}

\message{labels,}

\let\@@labeldef=\xdef
\newif\if@labelfile
\newwrite\@labelfile
\let\@prelabellist=\empty

\def\label#1#2{\trim#1\to\@@labarg\edef\@@labtext{#2}%
     \edef\@@labname{lab@\@@labarg}%
     \useafter\ifundefined\@@labname\then\else\@yeslab\fi
     \useafter\@@labeldef\@@labname{#2}%
     \ifstreaming
          \expandafter\toks@\expandafter\expandafter\expandafter
               {\csname\@@labname\endcsname}%
          \immediate\write\streamout{\noexpand\label{\@@labarg}{\the\toks@}}%
     \fi}
\def\@yeslab{%
     \useafter\ifundefined{if\@@labname}\then
          \err@labelredef\@@labarg
     \else\useif{if\@@labname}\then
               \err@labelredef\@@labarg
          \else\global\usename{\@@labname true}%
               \useafter\ifundefined{pre\@@labname}\then
               \else\useafter\ifx{pre\@@labname}\@@labtext
                    \else\err@badlabelmatch\@@labarg
                    \fi
               \fi
               \if@labelfile
               \else\global\@labelfiletrue
                    \immediate\write\sixt@@n{--> Creating file \jobname.lab}%
                    \immediate\openout\@labelfile=\jobname.lab
               \fi
               \immediate\write\@labelfile
                    {\noexpand\prelabel{\@@labarg}{\@@labtext}}%
          \fi
     \fi}

\def\putlab#1{\trim#1\to\@@labarg\edef\@@labname{lab@\@@labarg}%
     \useafter\ifundefined\@@labname\then\@nolab\else\usename\@@labname\fi}
\def\@nolab{%
     \useafter\ifundefined{pre\@@labname}\then
          \undefinedlabelformat
          \err@needlabel\@@labarg
          \useafter\xdef\@@labname{\undefinedlabelformat}%
     \else\usename{pre\@@labname}%
          \useafter\xdef\@@labname{\usename{pre\@@labname}}%
     \fi
     \useafter\newif{if\@@labname}%
     \expandafter\additemR\@@labarg\to\@prelabellist}

\def\prelabel#1{\useafter\gdef{prelab@#1}}

\def\ifundefinedlabel#1\then{%
     \expandafter\ifx\csname lab@#1\endcsname\relax}
\def\useiflab#1\then{\csname iflab@#1\endcsname}

\def\prelabelcheck{{%
     \def\^^\##1{\useiflab{##1}\then\else\err@undefinedlabel{##1}\fi}%
     \@prelabellist}}

\message{equation numbering,}

\newcount\chapternum
\newcount\sectionnum
\newcount\subsectionnum
\newcount\equationnum
\newcount\subequationnum
\newcount\figurenum
\newcount\subfigurenum
\newcount\tablenum
\newcount\subtablenum

\newif\if@subeqncount
\newif\if@subfigcount
\newif\if@subtblcount

\def\newchapternum{\newsectionnum=\z@\@resetnum\chapternum}
\def\newsectionnum{\newsubsectionnum=\z@\@resetnum\sectionnum}
\def\newsubsectionnum{\newequationnum=\z@\newfigurenum=\z@\newtablenum=\z@
     \@resetnum\subsectionnum}
\def\newequationnum{\newsubequationnum=\z@\@resetnum\equationnum}
\def\newsubequationnum{\@resetnum\subequationnum}
\def\newfigurenum{\newsubfigurenum=\z@\@resetnum\figurenum}
\def\newsubfigurenum{\@resetnum\subfigurenum}
\def\newtablenum{\newsubtablenum=\z@\@resetnum\tablenum}
\def\newsubtablenum{\@resetnum\subtablenum}

\def\@resetnum#1{\global\advance#1by1 \edef\next{\the#1\relax}\global#1}

\newchapternum=0

\def\chapternumstyle#1{\@setnumstyle\chapternum{#1}}
\def\sectionnumstyle#1{\@setnumstyle\sectionnum{#1}}
\def\subsectionnumstyle#1{\@setnumstyle\subsectionnum{#1}}
\def\equationnumstyle#1{\@setnumstyle\equationnum{#1}}
\def\subequationnumstyle#1{\@setnumstyle\subequationnum{#1}%
     \ifblank\subequationnumstyle\then\global\@subeqncountfalse\fi
     \ignorespaces}
\def\figurenumstyle#1{\@setnumstyle\figurenum{#1}}
\def\subfigurenumstyle#1{\@setnumstyle\subfigurenum{#1}%
     \ifblank\subfigurenumstyle\then\global\@subfigcountfalse\fi
     \ignorespaces}
\def\tablenumstyle#1{\@setnumstyle\tablenum{#1}}
\def\subtablenumstyle#1{\@setnumstyle\subtablenum{#1}%
     \ifblank\subtablenumstyle\then\global\@subtblcountfalse\fi
     \ignorespaces}

\def\eqnlabel#1{%
     \if@subeqncount
          \newsubequationnum=\next
     \else\newequationnum=\next
          \ifblank\subequationnumstyle\then
          \else\global\@subeqncounttrue
               \newsubequationnum=\@ne
          \fi
     \fi
     \label{#1}{\puteqnformat}(\puteqn{#1})%
     \ifdraft\rlap{\hskip.1in{\tt#1}}\fi}

\let\puteqn=\putlab

\def\equation#1#2{\useafter\gdef{eqn@#1}{#2\eqno\eqnlabel{#1}}}
\def\Equation#1{\useafter\gdef{eqn@#1}}

\def\putequation#1{\useafter\ifundefined{eqn@#1}\then
     \err@undefinedeqn{#1}\else\usename{eqn@#1}\fi}

\def\eqnseriesstyle#1{\gdef\@eqnseriesstyle{#1}}
\def\begineqnseries{\subequationnumstyle{\@eqnseriesstyle}%
     \defaultoption[]\@begineqnseries}
\def\@begineqnseries[#1]{\edef\@@eqnname{#1}}
\def\endeqnseries{\subequationnumstyle{blank}%
     \expandafter\ifnoarg\@@eqnname\then
     \else\label\@@eqnname{\puteqnformat}%
     \fi
     \aftergroup\ignorespaces}

\def\figlabel#1{%
     \if@subfigcount
          \newsubfigurenum=\next
     \else\newfigurenum=\next
          \ifblank\subfigurenumstyle\then
          \else\global\@subfigcounttrue
               \newsubfigurenum=\@ne
          \fi
     \fi
     \label{#1}{\putfigformat}\putfig{#1}%
     {\def\marginnoteformat{\tt}\marginnote{#1}}}

\let\putfig=\putlab

\def\figseriesstyle#1{\gdef\@figseriesstyle{#1}}
\def\beginfigseries{\subfigurenumstyle{\@figseriesstyle}%
     \defaultoption[]\@beginfigseries}
\def\@beginfigseries[#1]{\edef\@@figname{#1}}
\def\endfigseries{\subfigurenumstyle{blank}%
     \expandafter\ifnoarg\@@figname\then
     \else\label\@@figname{\putfigformat}%
     \fi
     \aftergroup\ignorespaces}

\def\tbllabel#1{%
     \if@subtblcount
          \newsubtablenum=\next
     \else\newtablenum=\next
          \ifblank\subtablenumstyle\then
          \else\global\@subtblcounttrue
               \newsubtablenum=\@ne
          \fi
     \fi
     \label{#1}{\puttblformat}\puttbl{#1}%
     {\def\marginnoteformat{\tt}\marginnote{#1}}}

\let\puttbl=\putlab

\def\tblseriesstyle#1{\gdef\@tblseriesstyle{#1}}
\def\begintblseries{\subtablenumstyle{\@tblseriesstyle}%
     \defaultoption[]\@begintblseries}
\def\@begintblseries[#1]{\edef\@@tblname{#1}}
\def\endtblseries{\subtablenumstyle{blank}%
     \expandafter\ifnoarg\@@tblname\then
     \else\label\@@tblname{\puttblformat}%
     \fi
     \aftergroup\ignorespaces}

\message{reference numbering,}

\newcount\referencenum \referencenum=0
\newcount\@@prerefcount \@@prerefcount=0
\newcount\@@thisref
\newcount\@@lastref
\newcount\@@loopref
\newcount\@@refseq
\newdimen\refnumindent
\let\@undefreflist=\empty

\def\referencenumstyle#1{\@setnumstyle\referencenum{#1}}

\def\referencestyle#1{\usename{@ref#1}}

\def\@refsequential{%
     \gdef\@refpredef##1{\global\advance\referencenum by\@ne
          \let\^^\=0\label{##1}{\^^\{\the\referencenum}}%
          \useafter\gdef{ref@\the\referencenum}{{##1}{\undefinedlabelformat}}}%
     \gdef\@reference##1##2{%
          \ifundefinedlabel##1\then
          \else\def\^^\####1{\global\@@thisref=####1\relax}\putlab{##1}%
               \useafter\gdef{ref@\the\@@thisref}{{##1}{##2}}%
          \fi}%
     \gdef\endputreferences{%
          \loop\ifnum\@@loopref<\referencenum
                    \advance\@@loopref by\@ne
                    \expandafter\expandafter\expandafter\@printreference
                         \csname ref@\the\@@loopref\endcsname
          \repeat
          \par}}

\def\@refpreordered{%
     \gdef\@refpredef##1{\global\advance\referencenum by\@ne
          \additemR##1\to\@undefreflist}%
     \gdef\@reference##1##2{%
          \ifundefinedlabel##1\then
          \else\global\advance\@@loopref by\@ne
               {\let\^^\=0\label{##1}{\^^\{\the\@@loopref}}}%
               \@printreference{##1}{##2}%
          \fi}
     \gdef\endputreferences{%
          \def\^^\####1{\useiflab{####1}\then
               \else\reference{####1}{\undefinedlabelformat}\fi}%
          \@undefreflist
          \par}}

\def\beginprereferences{\par
     \def\reference##1##2{\global\advance\referencenum by1\@ne
          \let\^^\=0\label{##1}{\^^\{\the\referencenum}}%
          \useafter\gdef{ref@\the\referencenum}{{##1}{##2}}}}
\def\endprereferences{\global\@@prerefcount=\the\referencenum\par}

\def\beginputreferences{\par
     \refnumindent=\z@\@@loopref=\z@
     \loop\ifnum\@@loopref<\referencenum
               \advance\@@loopref by\@ne
               \setbox\z@=\hbox{\referencenum=\@@loopref
                    \referencenumformat\enskip}%
               \ifdim\wd\z@>\refnumindent\refnumindent=\wd\z@\fi
     \repeat
     \putreferenceformat
     \@@loopref=\z@
     \loop\ifnum\@@loopref<\@@prerefcount
               \advance\@@loopref by\@ne
               \expandafter\expandafter\expandafter\@printreference
                    \csname ref@\the\@@loopref\endcsname
     \repeat
     \let\reference=\@reference}

\def\@printreference#1#2{\ifx#2\undefinedlabelformat\err@undefinedref{#1}\fi
     \noindent\ifdraft\rlap{\hskip\hsize\hskip.1in \tt#1}\fi
     \llap{\referencenum=\@@loopref\referencenumformat\enskip}#2\par}

\def\reference#1#2{{\par\refnumindent=\z@\putreferenceformat\noindent#2\par}}

\def\putref#1{\trim#1\to\@@refarg
     \expandafter\ifnoarg\@@refarg\then
          \toks@={\relax}%
     \else\@@lastref=-\@m\def\@@refsep{}\def\@more{\@nextref}%
          \toks@={\@nextref#1,,}%
     \fi\the\toks@}
\def\@nextref#1,{\trim#1\to\@@refarg
     \expandafter\ifnoarg\@@refarg\then
          \let\@more=\relax
     \else\ifundefinedlabel\@@refarg\then
               \expandafter\@refpredef\expandafter{\@@refarg}%
          \fi
          \def\^^\##1{\global\@@thisref=##1\relax}%
          \global\@@thisref=\m@ne
          \setbox\z@=\hbox{\putlab\@@refarg}%
     \fi
     \advance\@@lastref by\@ne
     \ifnum\@@lastref=\@@thisref\advance\@@refseq by\@ne\else\@@refseq=\@ne\fi
     \ifnum\@@lastref<\z@
     \else\ifnum\@@refseq<\thr@@
               \@@refsep\def\@@refsep{,}%
               \ifnum\@@lastref>\z@
                    \advance\@@lastref by\m@ne
                    {\referencenum=\@@lastref\putrefformat}%
               \else\undefinedlabelformat
               \fi
          \else\def\@@refsep{--}%
          \fi
     \fi
     \@@lastref=\@@thisref
     \@more}

\message{streaming,}

\newif\ifstreaming

\def\streamto{\defaultoption[\jobname]\@streamto}
\def\@streamto[#1]{\global\streamingtrue
     \immediate\write\sixt@@n{--> Streaming to #1.str}%
     \newwrite\streamout\immediate\openout\streamout=#1.str }

\def\streamfrom{\defaultoption[\jobname]\@streamfrom}
\def\@streamfrom[#1]{\newread\streamin\openin\streamin=#1.str
     \ifeof\streamin
          \expandafter\err@nostream\expandafter{#1.str}%
     \else\immediate\write\sixt@@n{--> Streaming from #1.str}%
          \let\@@labeldef=\gdef
          \ifstreaming
               \edef\@elc{\endlinechar=\the\endlinechar}%
               \endlinechar=\m@ne
               \loop\read\streamin to\@@scratcha
                    \ifeof\streamin
                         \streamingfalse
                    \else\toks@=\expandafter{\@@scratcha}%
                         \immediate\write\streamout{\the\toks@}%
                    \fi
                    \ifstreaming
               \repeat
               \@elc
               \input #1.str
               \streamingtrue
          \else\input #1.str
          \fi
          \let\@@labeldef=\xdef
     \fi}

\def\streamcheck{\ifstreaming
     \immediate\write\streamout{\pagenum=\the\pagenum}%
     \immediate\write\streamout{\footnotenum=\the\footnotenum}%
     \immediate\write\streamout{\referencenum=\the\referencenum}%
     \immediate\write\streamout{\chapternum=\the\chapternum}%
     \immediate\write\streamout{\sectionnum=\the\sectionnum}%
     \immediate\write\streamout{\subsectionnum=\the\subsectionnum}%
     \immediate\write\streamout{\equationnum=\the\equationnum}%
     \immediate\write\streamout{\subequationnum=\the\subequationnum}%
     \immediate\write\streamout{\figurenum=\the\figurenum}%
     \immediate\write\streamout{\subfigurenum=\the\subfigurenum}%
     \immediate\write\streamout{\tablenum=\the\tablenum}%
     \immediate\write\streamout{\subtablenum=\the\subtablenum}%
     \immediate\closeout\streamout
     \fi}


\def\err@badtypesize{%
     \errhelp={The limited availability of certain fonts requires^^J%
          that the base type size be 10pt, 12pt, or 14pt.^^J}%
     \errmessage{--> Illegal base type size}}

\def\err@badsizechange{\immediate\write\sixt@@n
     {--> Size change not allowed in math mode, ignored}}

\def\err@sizetoolarge#1{\immediate\write\sixt@@n
     {--> \noexpand#1 too big, substituting HUGE}}

\def\err@sizenotavailable#1{\immediate\write\sixt@@n
     {--> Size not available, \noexpand#1 ignored}}

\def\err@fontnotavailable#1{\immediate\write\sixt@@n
     {--> Font not available, \noexpand#1 ignored}}

\def\err@sltoit{\immediate\write\sixt@@n
     {--> Style \noexpand\sl not available, substituting \noexpand\it}%
     \it}

\def\err@bfstobf{\immediate\write\sixt@@n
     {--> Style \noexpand\bfs not available, substituting \noexpand\bf}%
     \bf}

\def\err@badgroup#1#2{%
     \errhelp={The block you have just tried to close was not the one^^J%
          most recently opened.^^J}%
     \errmessage{--> \noexpand\end{#1} doesn't match \noexpand\begin{#2}}}

\def\err@badcountervalue#1{\immediate\write\sixt@@n
     {--> Counter (#1) out of bounds}}

\def\err@extrafootnotemark{\immediate\write\sixt@@n
     {--> \noexpand\footnotemark command
          has no corresponding \noexpand\footnotetext}}

\def\err@extrafootnotetext{%
     \errhelp{You have given a \noexpand\footnotetext command without first
          specifying^^Ja \noexpand\footnotemark.^^J}%
     \errmessage{--> \noexpand\footnotetext command has no corresponding
          \noexpand\footnotemark}}

\def\err@labelredef#1{\immediate\write\sixt@@n
     {--> Label "#1" redefined}}

\def\err@badlabelmatch#1{\immediate\write\sixt@@n
     {--> Definition of label "#1" doesn't match value in \jobname.lab}}

\def\err@needlabel#1{\immediate\write\sixt@@n
     {--> Label "#1" cited before its definition}}

\def\err@undefinedlabel#1{\immediate\write\sixt@@n
     {--> Label "#1" cited but never defined}}

\def\err@undefinedeqn#1{\immediate\write\sixt@@n
     {--> Equation "#1" not defined}}

\def\err@undefinedref#1{\immediate\write\sixt@@n
     {--> Reference "#1" not defined}}

\def\err@nostream#1{%
     \errhelp={You have tried to input a stream file that doesn't exist.^^J}%
     \errmessage{--> Stream file #1 not found}}

\message{jyTeX initialization}

\everyjob{\immediate\write16{--> jyTeX version \fmtversion}%
     \edef\@@jobname{\jobname}%
     \edef\jobname{\@@jobname}%
     \settime
     \openin0=\jobname.lab
     \ifeof0
     \else\closein0
          \immediate\write16{--> Getting labels from file \jobname.lab}%
          \input\jobname.lab
     \fi}


\def\fixedskipslist{%
     \^^\{\topskip}%
     \^^\{\splittopskip}%
     \^^\{\maxdepth}%
     \^^\{\skip\topins}%
     \^^\{\skip\footins}%
     \^^\{\headskip}%
     \^^\{\footskip}}

\def\scalingskipslist{%
     \^^\{\p@renwd}%
     \^^\{\delimitershortfall}%
     \^^\{\nulldelimiterspace}%
     \^^\{\scriptspace}%
     \^^\{\jot}%
     \^^\{\normalbaselineskip}%
     \^^\{\normallineskip}%
     \^^\{\normallineskiplimit}%
     \^^\{\baselineskip}%
     \^^\{\lineskip}%
     \^^\{\lineskiplimit}%
     \^^\{\bigskipamount}%
     \^^\{\medskipamount}%
     \^^\{\smallskipamount}%
     \^^\{\parskip}%
     \^^\{\parindent}%
     \^^\{\abovedisplayskip}%
     \^^\{\belowdisplayskip}%
     \^^\{\abovedisplayshortskip}%
     \^^\{\belowdisplayshortskip}%
     \^^\{\abovechapterskip}%
     \^^\{\belowchapterskip}%
     \^^\{\abovesectionskip}%
     \^^\{\belowsectionskip}%
     \^^\{\abovesubsectionskip}%
     \^^\{\belowsubsectionskip}}


\def\twoupsetup{
     \topmargin=.75in
     \leftmargin=.5in
     \vsize=6.9in
     \hsize=4.75in
     \fullhsize=10in
     \let\draft=\relax}

\outputstyle{normal}                             

\def\marginnoteformat{\subscriptsize             
     \hsize=1in \baselinestretch=1000 \everypar={}%
     \tolerance=5000 \hbadness=5000 \parskip=0pt \parindent=0pt
     \leftskip=0pt \rightskip=0pt \raggedright}

\head={\ifdraft\normalfonts\it\hfil DRAFT\hfil   
     \llap{\number\day\ \monthword\month\ \militarytime}\else\hfil\fi}
\foot={\hfil\normalfonts\numstyle\pagenum\hfil}  

\normalbaselineskip=12pt                         
\normallineskip=0pt                              
\normallineskiplimit=0pt                         
\normalbaselines                                 

\topskip=.85\baselineskip \splittopskip=\topskip \headskip=2\baselineskip
\footskip=\headskip

\pagenumstyle{arabic}                            

\parskip=0pt                                     
\parindent=20pt                                  

\baselinestretch=1000                            


\chapterstyle{left}                              
\chapternumstyle{blank}                          
\def\chapterbreak{\newpage}                      
\abovechapterskip=0pt                            
\belowchapterskip=1.5\baselineskip               
     plus.38\baselineskip minus.38\baselineskip
\def\chapternumformat{\numstyle\chapternum.}     

\sectionstyle{left}                              
\sectionnumstyle{blank}                          
\def\sectionbreak{\vskip0pt plus4\baselineskip\penalty-100
     \vskip0pt plus-4\baselineskip}              
\abovesectionskip=1.5\baselineskip               
     plus.38\baselineskip minus.38\baselineskip
\belowsectionskip=\the\baselineskip              
     plus.25\baselineskip minus.25\baselineskip
\def\sectionnumformat{
     \ifblank\chapternumstyle\then\else\numstyle\chapternum.\fi
     \numstyle\sectionnum.}

\subsectionstyle{left}                           
\subsectionnumstyle{blank}                       
\def\subsectionbreak{\vskip0pt plus4\baselineskip\penalty-100
     \vskip0pt plus-4\baselineskip}              
\abovesubsectionskip=\the\baselineskip           
     plus.25\baselineskip minus.25\baselineskip
\belowsubsectionskip=.75\baselineskip            
     plus.19\baselineskip minus.19\baselineskip
\def\subsectionnumformat{
     \ifblank\chapternumstyle\then\else\numstyle\chapternum.\fi
     \ifblank\sectionnumstyle\then\else\numstyle\sectionnum.\fi
     \numstyle\subsectionnum.}


\footnotenumstyle{symbols}                       
\footnoteskip=0pt                                
\def\footnotenumformat{\numstyle\footnotenum}    
\def\footnoteformat{\footnotesize                
     \everypar={}\parskip=0pt \parfillskip=0pt plus1fil
     \leftskip=1em \rightskip=0pt
     \spaceskip=0pt \xspaceskip=0pt
     \def\\{\ifhmode\ifnum\lastpenalty=-10000
          \else\hfil\penalty-10000 \fi\fi\ignorespaces}}


\def\undefinedlabelformat{$\bullet$}             


\equationnumstyle{arabic}                        
\subequationnumstyle{blank}                      
\figurenumstyle{arabic}                          
\subfigurenumstyle{blank}                        
\tablenumstyle{arabic}                           
\subtablenumstyle{blank}                         

\eqnseriesstyle{alphabetic}                      
\figseriesstyle{alphabetic}                      
\tblseriesstyle{alphabetic}                      

\def\puteqnformat{\hbox{
     \ifblank\chapternumstyle\then\else\numstyle\chapternum.\fi
     \ifblank\sectionnumstyle\then\else\numstyle\sectionnum.\fi
     \ifblank\subsectionnumstyle\then\else\numstyle\subsectionnum.\fi
     \numstyle\equationnum
     \numstyle\subequationnum}}
\def\putfigformat{\hbox{
     \ifblank\chapternumstyle\then\else\numstyle\chapternum.\fi
     \ifblank\sectionnumstyle\then\else\numstyle\sectionnum.\fi
     \ifblank\subsectionnumstyle\then\else\numstyle\subsectionnum.\fi
     \numstyle\figurenum
     \numstyle\subfigurenum}}
\def\puttblformat{\hbox{
     \ifblank\chapternumstyle\then\else\numstyle\chapternum.\fi
     \ifblank\sectionnumstyle\then\else\numstyle\sectionnum.\fi
     \ifblank\subsectionnumstyle\then\else\numstyle\subsectionnum.\fi
     \numstyle\tablenum
     \numstyle\subtablenum}}


\referencestyle{sequential}                      
\referencenumstyle{arabic}                       
\def\putrefformat{\numstyle\referencenum}        
\def\referencenumformat{\numstyle\referencenum.} 
\def\putreferenceformat{
     \everypar={\hangindent=1em \hangafter=1 }%
     \def\\{\hfil\break\null\hskip-1em \ignorespaces}%
     \leftskip=\refnumindent\parindent=0pt \interlinepenalty=1000 }


\normalsize


\def\fmtversion{2.6M (June 1992)}

\catcode`\@=12

\typesize=10pt \magnification=1200 \baselineskip17truept
\footnotenumstyle{arabic} \hsize=6truein\vsize=8.5truein

\sectionnumstyle{blank}
\chapternumstyle{blank}
\chapternum=1
\sectionnum=1
\pagenum=0

\def\begintitle{\pagenumstyle{blank}\parindent=0pt
\begin{narrow}[0.4in]}
\def\endtitle{\end{narrow}\newpage\pagenumstyle{arabic}}


\def\beginexercise{\vskip 20truept\parindent=0pt\begin{narrow}[10
truept]}
\def\endexercise{\vskip 10truept\end{narrow}}


\def\eql#1{\eqno\eqnlabel{#1}}
\def\ref{\reference}
\def\peq{\puteqn}
\def\pref{\putref}

\def\mgn{\marginnote}
\def\bex{\begin{exercise}}
\def\eex{\end{exercise}}



\font\goth=eufm10  

\def\StretchRtArr#1{{\count255=0\loop\relbar\joinrel\advance\count255 by1
\ifnum\count255<#1\repeat\rightarrow}}
\def\StretchLtArr#1{\,{\leftarrow\!\!\count255=0\loop\relbar
\joinrel\advance\count255 by1\ifnum\count255<#1\repeat}}

\def\StretchLRtArr#1{\,{\leftarrow\!\!\count255=0\loop\relbar\joinrel\advance
\count255 by1\ifnum\count255<#1\repeat\rightarrow\,\,}}

\def\mbox#1{{\leavevmode\hbox{#1}}}

\def\hspace#1{{\phantom{\mbox#1}}}

\def\gS{\mbox{{\goth\char83}}}
\def\gF{\mbox{{\goth\char70}}}

\def\al{\alpha}
\def\bom{{\bmit\omega}}
\def\be{\beta}

\def\de{\delta}
\def\Ga{\Gamma}

\def\si{\sigma}

\def\ze{\zeta}

\def\caN{{\cal N}}

\def\caE{{\cal E}}
\def\caF{{\cal F}}

\def\caR{{\cal R}}

\def\caB{{\cal B}}

\def\sc{{\rm sc }}

\def\zf{$\zeta$--function}
\def\zfs{$\zeta$--functions}


\def\frac#1/#2{\leavevmode\kern.1em
\raise.5ex\hbox{\the\scriptfont0 #1}\kern-.1em/\kern-.15em
\lower.25ex\hbox{\the\scriptfont0 #2}}
\def\sfrac#1/#2{\leavevmode\kern.1em
\raise.5ex\hbox{\the\scriptscriptfont0 #1}\kern-.1em/\kern-.15em
\lower.25ex\hbox{\the\scriptscriptfont0 #2}}

\def\gtorder{\mathrel{\raise.3ex\hbox{$>$}\mkern-14mu
             \lower0.6ex\hbox{$\sim$}}}
\def\ltorder{\mathrel{\raise.3ex\hbox{$<$}\mkern-14mu
             \lower0.6ex\hbox{$\sim$}}}

\def\semidirprod{\rlap{\ss C}\raise1pt\hbox{$\mkern.75mu\times$}}
\def\for{\lower6pt\hbox{$\Big|$}}
\def\fish{\kern-.25em{\phantom{abcde}\over \phantom{abcde}}\kern-.25em}


\def\boxit#1{\vbox{\hrule\hbox{\vrule\kern3pt
        \vbox{\kern3pt#1\kern3pt}\kern3pt\vrule}\hrule}}
\def\dalemb#1#2{{\vbox{\hrule height .#2pt
        \hbox{\vrule width.#2pt height#1pt \kern#1pt \vrule
                width.#2pt} \hrule height.#2pt}}}

\def\frac#1#2{{{#1}\over{#2}}}

\def\noin{\noindent}

\def\comb#1#2{{\left(#1\atop#2\right)}}

\def\etc{{\it etc. }}

\def\eg{{\it e.g.}}
\def\ie{{\it i.e. }}

\def\pa{\partial}



\def\3j#1#2#3#4#5#6{\left\lgroup\matrix{#1&#2&#3\cr#4&#5&#6\cr}
\right\rgroup}

\def\man{{\cal M}}

\def\m?{\mgn{?}}

\def\pa{\partial}

\def\beq{\begin{eqnarray}}
\def\eeq{\end{eqnarray}}


\def\cmp#1#2#3{{\it Comm. Math. Phys.} {\bf {#1}} ({#2}) #3}
\def\cqg#1#2#3{{\it Class. Quant. Grav.} {\bf {#1}} ({#2}) #3}

\def\jgp#1#2#3{{\it J. Geom. and Phys.} {\bf {#1}} ({#2}) #3}

\def\jpa#1#2#3{{\it J. Phys.} {\bf A{#1}} ({#2}) #3}

\def\np#1#2#3{{\it Nucl. Phys.} {\bf B{#1}} ({#2}) #3}

\def\pl#1#2#3{{\it Phys. Lett.} {\bf {#1}} ({#2}) #3}

\def\prD#1#2#3{{\it Phys. Rev.} {\bf D{#1}} ({#2}) #3}

\def\am#1#2#3{{\it Acta Mathematica} {\bf {#1}} ({#2}) #3}

\def\dmj#1#2#3{{\it Duke Math. J.} {\bf {#1}} ({#2}) #3}

\def\jpamt#1#2#3{{\it J. Phys.A:Math.Theor.} {\bf{#1}} ({#2}) #3}

\def\ma#1#2#3{{\it Math. Ann.} {\bf {#1}} ({#2}) #3}

\def\plb#1#2#3{{\it Phys. Letts.} {\bf {B#1}} ({#2}) #3}

\begin{title}
\vglue 0.5truein
\vskip15truept
\centertext {\Bigfonts \bf  Conformal anomalies for higher derivative} \vskip7truept
\vskip10truept\centertext{\Bigfonts \bf free critical $p$--forms on even spheres}
 \vskip7truept
\vskip10truept\centertext{\Bigfonts \bf }
 \vskip 20truept
\centertext{J.S.Dowker\footnote{dowkeruk@yahoo.co.uk}} \vskip 7truept \centertext{\it
Theory Group,} \centertext{\it School of Physics and Astronomy,} \centertext{\it The
University of Manchester,} \centertext{\it Manchester, England} \vskip 7truept
\centertext{}

\vskip 5truept \vskip 0truept
\begin{narrow}
The conformal anomaly is computed on even $d$--spheres for a $p$--form propagating
according to  the Branson--Gover higher derivative, conformally covariant operators. The
system is set up on a $q$--deformed sphere and the conformal anomaly is computed as a
rational function of the derivative order, $2k$, and of $q$. The anomaly is shown to be an
extremum at the round sphere ($q=1$) only for $k<d/2$. At these integer values,
therefore, the entanglement entropy is minus the conformal anomaly, as usual.

The unconstrained $p$--form conformal anomaly on the full sphere is shown to be given by
an integral over the Plancherel measure for a coexact form on hyperbolic space in one
dimension higher.

A natural ghost sum is constructed and leads to quantities which, for critical forms, \ie
when $2k=d-2p$, are, remarkably, a simple combination of standard quantities, for usual
second order, $k=1$, propagation, when these are available. Our values coincide with a
recent hyperbolic computation of David and Mukherjee.

Values are suggested for the Casimir energy on the Einstein cylinder  from the behaviour
of the conformal anomaly as $q\to0$ and compared with known results written as
alternating sums over scalar values.

\end{narrow}
\vskip 5truept
\vskip 60truept
\vfil
\end{title}
\pagenum=0
\newpage

\section{\bf 1. Introduction}

In earlier works, [\pref{dowgjmsren}], I discussed the construction of the (one loop)
effective actions for higher derivative scalars and spinors using the spherical product forms
of the GJMS--type propagation operators (kinetic operators) for all even sphere
dimensions.

Working in even dimensions, the interest was in calculating the coefficient of the
logarithmic (or divergent) term in the effective action (`free energy'). Technically this
amounted to evaluating the relevant \zf\ at 0.

For $p$--forms at that time, [\pref{dowCTpform}], I was able only to discuss second order
operators owing to the lack of ( knowledge of) a conformal higher derivative GJMS--type
operator. I now propose to take the Branson--Gover operators as the appropriate ones.
The genesis of these operators lies in papers by Branson, [\pref{Branson}], and developed
by Branson and Gover, [\pref{BandG}], but for convenience I will refer to Fischmann and
Somberg, [\pref{FandS}],  who have a combinatorial approach and a useful set of
references.

The mathematical motivation comes from conformal differential geometry which has been
subject to very substantial development. I will, therefore, simply posit the relevant
formulae and proceed to a workaday calculation of the `conformal anomaly'. I would also
wish to discuss the formal functional determinant but I leave this for another time.

I calculate coexact form quantities and combine them to give the unconstrained form ones.
These I assemble into  a `gauge invariant' combination, although I give no detailed field
theoretic justification of this.

I will, throughout, refer to a conformal anomaly and an effective action as shorthand
terminology for $\sim\ze(0)$ and $\sim\ze'(0)$ of the field in question.

I work on a $q$ conically deformed sphere which is the $d$--dimensional periodic spherical
$q$--lune described  in [\pref{dowgjmsren}] to where I refer for basic explanation.

\section{\bf2. The operators}
The construction of the Branson--Gover higher--derivative, conformally covariant operators
on a general manifold is somewhat complicated and need not detain us. However, on
conformal Einstein manifolds they have the remarkable property of being factorisable. For
spheres, this has been known for a long time, [\pref{Branson}].

Perhaps the reason why the operators have not appeared much in the physics literature, is
that the $\de d$ and $d\de$ parts are weighted differently, which seems unnaturally
different to the de Rham operator but which is necessary for conformal covariance,
[\pref{Branson}].

Special amongst the operators are the `critical' ones, where the derivative order, $2k$, of
the propagation operator is related to the form order, $p$, and manifold dimension $d$ by
  $$
        2k=d-2p
  $$
so that when $k=1$ (the usual case), $p=d/2-1$, the conformal value. In odd dimensions
operators exist for all values of the derivative order.

The general operator factorises on Einstein manifolds, at least, but I consider here only
the sphere S$^d$ for even $d$. Although spectral resolutions exist for the other
manifolds, the sphere will be quite enough to begin with.

Several expressions are given in [\pref{FandS}] \footnote{ Note the ArXiv and published
versions differ in format.} in Theorems 4.3 and 4.4 corresponding to various restrictions on
the parameters. For example, defining $\be=d/2-p$, an expression is given in equn.(4.7)
(or (38)) in [\pref{FandS}] which, for even $d$, holds only in the non--supercritical case,
\ie\ $\be>k-1$.

I copy it here, choosing a unit $d$--sphere,
   $$
       L^{(p)}_{2k}=\be\prod_{j=0}^{k-1}\bigg[{\be+j+1\over \be+j}\de d
       +{\be-j-1\over\be-j}
       d\de +\big({\be}-j-1\big)\big({\be}+j+1\big)\bigg]\,.
       \eql{genop}
   $$

From this, one can extract the part which acts on the range of $\de, \caR(\de)$, \ie on
coexact forms,
       $$\eqalign{
           L^{CE(p)}_{2k}&=\big({\be}+k\big)\prod_{j=0}^{k-1}\bigg[\de d
           +\big(\be-{1\over2}\big)^2
           - (j+1/2)^2\bigg]\cr
      &=\big({\be}+k\big)\prod_{j=0}^{k-1}\big(B^2-(j+1/2)^2\big)
      \equiv\big({\be}+k\big) \prod_0^{k-1}\big(B^2-\al_j^2\big)\cr
   &=\big({\be}+k\big)\prod_0^{k-1} \big(B-\al_j\big)\big(B+\al_j\big)\,,
   \quad k=1,2,\ldots\,,
      }
  \eql{OmegaC}
  $$
where the pseudo--operator $B$ is defined by, \footnote{ $\de d$ is taken to be
positive.},
  $$
     B=\sqrt{\de d +\al^2(a,p)}\,,
  $$
with
    $$
         \al(a,p)=\be-1/2=a-p\,,\quad\quad a\equiv (d-1)/2\,.
    $$

Likewise, acting on $\caR(d)$, exact forms, the operator turns into
$$\eqalign{
          L^{E(p)}_{2k}&= \big({\be}-k\big)\prod_{j=0}^{k-1}\bigg[d\de
           +\big(\be+{1\over2}\big)^2 - \al_j^2\bigg]\cr
      &=\big({\be}-k\big)\prod_0^{k-1}\big(C^2-\al_j^2\big)\cr
   &=\big({\be}-k\big)\prod_0^{k-1} \big(C-\al_j\big)\big(C+\al_j\big)\,,
   \quad k=1,2,\ldots\,,
      }
  \eql{OmegaE}
  $$
with $ C=\sqrt{d\de+(\be+1/2)^2} $.

These expressions are given in Branson, [\pref{Branson}], Remark (3.30) and I will refer
to them as the Branson operators.

When $\be=k$, the exact part disappears {\it algebraically} from (\peq{genop}) leaving
just the coexact one. Since we are just at the lower limit of the condition $\be>k-1$, the
operators have been termed `critical',
      $$
               L^{(p)}_{d-2p}=L^{CE(p)}_{d-2p}\,,
      $$
and could be considered as  generalisations of the four dimensional Maxwell operator, $\de
d$. The critical exact operator is the Hodge star dual of the coexact one ($p\to d-p$).

Critical operators possess important factorisations involving the $Q$ curvature and the
gauge companion, $G$. It is easily seen from (\peq{OmegaC}) that
  $$\eqalign{
           L^{(p)}_{2k}&=2k\,\de\circ\prod_{j=0}^{k-2}\bigg[d\de
           +\big(\be-{1\over2}\big)^2- (j+1/2)^2\bigg]\circ d\,,\quad 2k=d-2p\cr
           &\equiv \de\circ Q^{(p+1)}_{k-2}\circ d\cr
           &\equiv \de \circ G^{(p+1)}_{k-1}\,.
      }
  \eql{Fact}
  $$
I do not use these

When restricted to $\caR(d)$ or to $\caR(\de)$ the remaining operators in Theorem 4.4
also reduce to the Branson operators, (\peq{OmegaC}) and (\peq{OmegaE}) which can
therefore be taken as valid for {\it all}  $\be$.

In addition there is the further condition that $k$ must be less than $d/2$. This is the
point at which a zero eigenvalue first appears (for even $d$), \ie\ the same limitation as
in the {\it scalar} GJMS case. This is because the eigenvalues are independent of the form
order, $p$. (Only the degeneracy depends on $p$.) For this reason operators with $k=d/2$
could also be deemed even more `critical'. For odd $d$, $k$ can be continued beyond
$d/2$.

Most mathematical activity is concerned with even dimensional manifolds and I
concentrate on these here too.

\section{\bf 3. The geometry}

The simplest course would be to take the full sphere as the manifold but, in order
ultimately to discuss R\'enyi and entanglement entropies, I will, as in the earlier works,
take the $q$--deformed sphere. More precisely, I take a $q$--lune which doubles up to
make a periodic $q$--deformed sphere. The complete mode set is then obtained by
combining those for absolute and relative boundary conditions on the boundary of the
lune.

When $q=1$, the lune is a hemisphere and the doubling gives the full round sphere.

\section{\bf 4. The calculation}

The immediate aim is to calculate various coexact spectral invariants for the critical
operators. I will embed these in the more general coexact ones, (\peq{OmegaC}), so that
the analysis in [\pref{dowCTpform}] can be drawn upon.

I will assume that, because of the Hodge decomposition, $\caR(\de)\oplus\caR(d)\oplus
\caN(d)\cap\caN(\de)$, the coexact quantities are spectrally sufficient. Then, for example,
the total $p$--form conformal anomaly is the combination,\footnote{ This combination
would ensure Hodge star duality in $d$ dimensions for the anomaly. For the zeta function,
the scaling coefficients would have to be inserted.}
  $$
        \ze_{tot}(0,p)=\ze^{CE}(0,p)+\ze^{CE}(0,p-1)\,,
        \eql{totcomb}
        $$
aince one still has,
   $$
       \ze^{E}(0,p)=\ze^{CE}(0,p-1)\,,
  $$
although there would be scaling considerations for the determinant.

It should be noted that the system here analysed is not quite the same as that in
[\pref{dowCTpform}] for which the relevant operator  is $\de d$ for all allowed dimensions
and form orders. The second order Branson operator, $ \sim \de d +\al^2-1/4 $, depends
explicitly on the form order and manifold  dimension. It could be looked upon as a
`conformally improved' de Rham Laplacian because, for scalars (0--forms), it is the
Penrose--Yamabe operator, while $\de d$ is the minimal one. There is, however,
agreement for critical forms, and their duals.

Since I am presently interested only in the conformal anomaly the overall normalisation,
$(\be+k)$, of the operator (\peq{OmegaC}) plays no role and I will ignore it.\footnote{
This would not strictly be so for the functional determinants but can be allowed for as a
change of scale involving the conformal anomaly and hence absorbed by renormalisation.}

The computation is given in [\pref{dowCTpform, dowgjmsren}] but I develope a few
essentials here.\footnote{  The general trend and much of the detail of the calculation are
valid for even and odd dimensions but I break up the analysis and just consider even
dimensions here.} The important fact for the furtherance of the analysis is that $B$ and
$C$  have linear eigenvalues,
  $$
       a +1+m\,,\quad m=0,1,2,\ldots \,,
  $$
which can usefully be organised into an associated \zf,
  $$
  \ze(s,a,p,q)\equiv\sum_{m=0}^{\infty} {d(m)\over (a+1+m)^s}\,,
  \eql{auxzf}
  $$
where $d(m)$ is the degeneracy of the sphere eigenlevel labelled by $m$ and depends on
the boundary conditions and $p$, $d$ and  $q$. The exact degeneracy can be obtained
from the coexact one by the replacement $p\to p-1$.

The degeneracy is best encoded in a generating function which acts as a square--root
`heat kernel' and allows the construction of the auxiliary \zf, (\peq{auxzf}), by Mellin
transform {\it directly}. The details are given in [\pref{dowCTpform}] and result in a linear
combination of Barnes \zfs, $\ze_d$,
$$\eqalign{
  \ze^{CE}_a(s,a,p,q)
&=(-1)^{p+1}\!\!\sum_{r=p+1}^d(-1)^r\bigg[ \comb{d-1}r\ze_d(s,a-p+r\mid\bom)\cr
&\hspace{****************}+
\comb{d-1}{r-1}\,\ze_d(s,a-p+r+q-1\mid\bom)\bigg]\,,
  }
  \eql{zet4}
  $$
where the vector, $\bom$, stands for the $d$--dimensional set $\bom=(q,1,\ldots,1)$.

Equation (\peq{zet4}) is for absolute boundary conditions. Coexact duality gives the
relative case,
$$
\ze^{CE}_r(s,a,p,q)=\ze^{CE}_a(s,a,d-1-p,q)\,,
$$
which has to be added to the absolute expression in order to get the coexact $p$--form
value on the {\it periodic} (double) lune, which is the sphere when $q=1$.

One can also subtract them and expect to get a quantity defined on the lune boundary, a
($d-1)$--sphere.

The expressions for more general spherical factorings have been given earlier, under a
different name, in [\pref{dowpform1}] equns (23) and (24). In the case of the full sphere,
formulae and comments can also be found in [\pref{DandKii}].

The importance of $\ze(s,a)$ is that it is thus readily evaluated, and that the derivatives
of the full \zf\ of $L$ at $s=0$ can be given in terms of it, as I now briefly outline.

The eigenproblem for $L^{CE(p)}_{2k}$ is solved by that for $B$ and its \zf\ is given
by\footnote{ The factors of ${\be\pm k}$ in (\peq{OmegaC}), (\peq{OmegaE}) have now
been discarded.},
    $$
    Z(s,k,p,d)=\sum_{m=0}^\infty\prod_{j=0}^{k-1}{d(m)
    \over[(a+1+\al_j+m)(a+1-\al_j+m)]^s}\,.
    $$
A formal expansion, [\pref{Dowcmp}],  in the $\al_j$ allows one to find the value at zero
as the average
    $$
    Z(0,k,p,d)={1\over 2k}\sum_{j=0}^{k-1}\bigg( \ze(0,a+\al_j)+\ze(0,a-\al_j)\bigg)\,,
    \eql{confanom}
    $$
and the derivative at 0 as the `corrected' sum,
  $$
    Z'(0,k,p,d)=\sum_{j=0}^{k-1}\bigg(\ze'(0,a+\al_j)+\ze'(0,a-\al_j)\bigg)\,+\, M(k)\,,
    \eql{effact}
    $$
where the polynomial $M(k)$ could be termed a multiplicative anomaly, but is really part
and parcel of the evaluation.  \footnote{ I have simplified the notation a little. The \zf\ on
the right is that displayed in (\peq{zet4}) omitting `$p,q$'.}

These are my calculational expressions for the coexact `conformal anomaly' and `effective
action' (up to divergences) in even dimensions. Apart from $M$, they have the form of a
linear sum over spectral quantities associated with each linear factor in the product
operator $L^{CE (p)}_{2k}$, (\peq{OmegaC}).

Importantly, the GJMS sum over the $\al_j$ parameters in (\peq{OmegaC}) can be
transferred to the Barnes \zfs\ making up the auxiliary \zf, $\ze(s,a,p,q)$. It is instructive
and computationally quickening to make the effect of this geometric sum explicit at this
stage by the relation, [\pref{dowgjmsren}],

 $$\eqalign{
&\sum_{j=0}^{k-1}\big(\zeta_d(s,a+j+1/2\mid{\bom})
+\zeta_d(s,a-j-1/2\mid{\bom})\big)\cr
&\hspace{****}=\ze_{d+1}(s,a+1/2-k\mid \bom,1)-\ze_{d+1}(s,a+1/2+k\mid\bom,1)\,,
}
\eql{holog}
 $$
which converts the sum of $k$ $d$--dimensional quantities into the {\it difference} of two
$(d+1)$--dimensional quantities in a holographic fashion.

Note that on the left of (\peq{holog}), $k$ is an integer while the right allows an
extension off the integers. In particular, one can easily differentiate with respect to $k$.

\section{\bf 5. The conformal anomaly}
Doing the GJMS sum produces for the (absolute) coexact conformal anomaly on the
$q$--lune, (\peq{confanom}) using (\peq{zet4}),
$$\eqalign{
  {k}\,Z_a(0,k,p,d,q)&\equiv C_a(p,d,q,k)\cr
&={(-1)^{p+1}\over2}\!\!\sum_{r=p+1}^d(-1)^r\bigg[
\comb{d-1}r\ze_{d+1}(0,\al+1/2+k+r\mid\bom,1)\cr
&\hspace{****}+
\comb{d-1}{r-1}\,\ze_{d+1}(0,\al-1/2+k+r+q\mid\bom,1)\bigg]\cr
&\hspace{******}-(k\to -k)\cr
&={(-1)^{p+d}\over2(d+1)!q}\sum_{r=p+1}^d(-1)^r\!\bigg[
\comb{d-1}r \big[B^{(d+1)}_{d+1}(\al+1/2+k+r|\bom,1)\cr
&\hspace{*************}-B^{(d+1)}_{d+1}(\al+1/2-k+r\mid\bom,1)\big]\cr
&\hspace{**********}+
\comb{d-1}{r-1}\,\big[B^{(d+1)}_{d+1}(d/2+1+p-r-k\mid\bom,1)\cr
&\hspace{**************}-B^{(d+1}_{d+1}(d/2+1+p-r+k\mid\bom,1)\big]\bigg]\,,\cr
  }
  \eql{zet41}
  $$
in terms of generalised Bernoulli functions\footnote{ I have applied a transformation to
the arguments of the last two in order to simplify the $q$--dependence.}, which are easily
evaluated as polynomials in $q$ and $k$, and so can be extended off the integers.
Particular examples are given {in the next section.

\section{\bf 6. Some results}

The results are in much the same form as those displayed in [\pref{dowgjmsren}]

I firstly present some examples for the absolute coexact conformal anomaly on the single
lune, (\peq{zet41}), as bi--polynomials in $k$ and $q$ for given $p$ and $d$.
   $$\eqalign{
       C_a(1,4,q,k)&=-{k\over240 q}\big(q^4+10(1-k^2)q^2
       +40(k^2-6)q-6k^4+20k^2-11\big)\cr
       C_a(2,4,q,k)&=-{k\over240q}\big(q^4+10(1-k^2)q^2+40k^2q-6k^4+20k^2-11\big)\cr
       C_a(1,6,q,k)&={k\over6048q}\big(10q^6+35(3-2k^2)q^4+210(k^2-1)(k^2-4)q^2\cr
       &\hspace{***}-168(30k^4-5k^2-36)q+60k^6-630k^4+1680k^2-955\big)\cr
       C_a(2,6,q,k)&={k\over30240q}(10q^6+35(3-2k^2)q^4+210(k^2-1)(k^2-4)q^2\cr
       &\hspace{***}-252(3k^4-25k^2+12)q+60k^6-630k^4+1680k^2-955\big)\,.\cr
       }
       \eql{calunea}
   $$

The relative value can be obtained using Hodge coexact duality, that is by sending $p\to
d-1-p$. Added to the absolute value it gives the (coexact) conformal anomaly on the
doubled lune, \ie\ on the $q$--deformed sphere. Subtraction  yields,
    $$
             C_a(p,d,q,k)-C_r(p,d,q,k)=(-1)^{d/2}\,k\,.
    $$

That this is independent of $q$ reflects the fact that it is a quantity associated with the
boundary of the $q$--lune, which is a full $(d-1)$-sphere.

Although not immediately enlightening, the formulae (\peq{calunea}) do reveal the
important fact that the coefficient of $q$ vanishes when $k$ is an allowed value, $k<d/2$.
This is a consequence of conformal invariance. It is also the case for scalars.

On the full sphere some values for $C^{CE}=C_a+C_r$ are,
  $$\eqalign{
      C^{CE}(1,4,1,k)&={k\over60}(3k^4-25k^2+60)\cr
      C^{CE}(2,6,1.k)&={k\over3780}(15k^6-294k^4+1715k^2-3780)\cr
     C^{CE}(3,8,1,k)&= {k\over181440}(35k^8-1350k^6+17199k^4-86100k^2+181440)\,.
      }
      \eql{casph}
  $$

As a non trivial check, the corresponding results for the scalar $0$--form agree with those
in [\pref{dowgjmsren}].\footnote{ Allowance has then to be made for the zero scalar
modes which are `missing' from the coexact spectrum. This amounts to the addition of a
term $k$ to $C$ for all $q$. The factor of $k$ arises because the missing mode contributes
$1$ to $Z(0)\sim C/k$.}

I pass on to the entanglement entropy (universal part of), and the derivative with respect
to $q$ taken at the round limit $q=1$.  Higher derivatives in the complete field theory
would lead onto the central charges. I will later also re--express the sphere coexact
conformal anomalies, (\peq{casph}), more interestingly.

\section{\bf 7.  Entanglement entropy. Derivatives.}

At this point, it is convenient to introduce the R\'enyi entropy, $S_n$, defined generally
by,
$$
   S_n={nW(1)-W(1/n)\over1-n}\,,
   \eql{renyi}
  $$
where $W(q)$ here is the effective action on the periodic $q$--lune. $n=1/q$ is the
R\'enyi, or replica, index. $S_1$ is the entanglement entropy\footnote{ I define the
entanglement entropy by the replica trick.} and $S'_1$ determines the central charge,
$C_T$, [\pref{Perlmutter}].

In even dimensions, the  universal component of $S_n$ is obtained by substituting the
value $\ze(0)$ for the effective action, $W$, in (\peq{renyi}). $\ze(s)$ is the spectral \zf\
of the propagating operator on the conically deformed manifold.

\begin{ignore}

In the situation here, we have all the necessary quantities obtained in the previous
sections and a few examples of $\gS_q(p,d,k)$ follow from (\peq{renyi}) and
(\peq{zet41}),
  $$
     \eqalign{
     \gS_q(1,4,k)&={k\over120}\big(q^3+q^2-(10k^2-11)q+30k^2-109\big)\cr
     \gS_q(2,6,k)&={k\over15120}\big(10q^5+10q^4-5(14k^2-13)(q+1)q^2\cr
     &\hspace{*******}+5(42k^4-224k^2+191)q-546k^4+5180k^2-14165\big)
     }
     \eql{renyi2}
  $$

\end{ignore}

At the round sphere ($q=1$), the R\'enyi entropy equals the entanglement entropy and
this, as calculation shows, equals (minus) the conformal anomaly, $k\ze(0)=C$ but {\it
only} when $k$ is an integer not above the larger critical limit \ie\ $k\le d/2$.

An equivalent statement is that the conformal anomaly on the $q$--deformed sphere is an
extremum, as $q$ varies, at the full sphere only for these non--supercritical $k$ integers.
The same holds for scalar and Dirac fields, [\pref{dowgjmsren}]. I give a few details for
present circumstances.

The derivative with respect to $q$ can be calculated generally from (\peq{zet41}) or
individually from any particular expression.such as (\peq{calunea}). The coexact result at
the round sphere is found to be,
  $$
       {\pa\over\pa q} C^{CE}(p,d,q,k)\bigg|_{q=1}={2\over(d+1)!}\comb{d-1}
       p\big((d/2)^2-k^2\big)\,.
       \ldots k^2\,.
  $$

Sending $p\to p-1$ and adding gives the quantity for a free $p$--form (\peq{totcomb}),
  $$
       {\pa\over\pa q} C^{tot}(p,d,q,k)\bigg|_{q=1}={2\over(d+1)!}\comb dp
       \big((d/2)^2-k^2\big)\,.\ldots k^2\,.
  $$
These show an explicit vanishing only for the allowed values of $k$.

This property carries through to any quantity constructed linearly from coexact ones.

\section{\bf8. Alternative factorisation}

For $k$ integral, as it mostly has been so far, the product can be written as the ratio of
two Gamma functions, in a familiar way,
  $$
        L^{CE(p)}_{2k}=\big({\be}+k\big){\Ga(B+k+1/2)\over\Ga(B-k+1/2)}\,,
        \eql{gammar}
  $$
which permits continuation in $k$, in particular to $k$ a half integer when the expression
again factorises, this time into a pseudo--operator,
  $$
  L^{CE(p)}_{2k}=\big({\be}+k\big)B\,\prod_{h=1}^l\big(B^2-h^2\big)\,.
  \quad k\equiv l+1/2\,.
  $$

Lowest examples are $\sqrt{\de d}\,$, $\sqrt{\de d+1}\,\de d\,$, $ \sqrt{\de d+4}\,(\de
d+3)\de d$ and should be considered as analogous to boundary or Neumann--Dirichlet
operators.

Taking the critical condition $2k=d-2p$ seriously suggests that these pseudo--operators
are relevant for coexact forms in odd dimensions.

Both factorisations are economically combined in the product, [\pref{Branson}],
     $$
          \big({\be}+k\big)\prod_{m=1}^{2k}\big(B-k+m\big)\,,
     $$
where $k$ is either an integer or a half--integer.

The $\caR(d)$ operators can likewise be expressed
   $$
          \big({\be}-k\big)\prod_{m=1}^{2k}\big(C-k+m\big)\,.
     $$

By continuation, conformal anomalies computed from these pseudo--operators would agree
with those displayed above, such as (\peq{casph}). Furthermore, we can investigate how
the spectral quantities vary as $k$ varies, in particular the derivatives with respect to $k$,
an example of which follows.

\section{\bf9. Plancherel form of the sphere conformal anomaly.}

It is productive to consider the $k$--derivative \footnote{ Compare the AdS/CFT
calculations of Diaz and Dorn, [\pref{DandD}].} of the $C$ quantities, \eg,
  $$
      {\pa\over\pa k}C^{tot}(p,d,q,k)\,,
  $$
for the complete (free) $p$--form, (\peq{totcomb}), on the full sphere.

A slightly involved calculation reveals that this has a product structure which can then be
integrated to yield the more congenial expression, ($\be=d/2-p$),
  $$
       C^{tot}(p,d,1,k)={2(-1)^{d/2}\over d!}\comb d p\int_0^kdt\,{1\over \be^2
       -t^2}\prod_{i=0}^{d/2}(i^2-t^2)\,,
       \eql{planch}
  $$
which again exhibits the correct dimension of a free $p$--form and Hodge duality (under
which, $\beta\to -\beta$). It agrees with known expressions for $0$--forms,
[\pref{Diaz,DowGJMS}].

The integrand is (proportional to) the continuation of the {\it coexact} Plancherel measure
for $p$--forms on H$^{d+1}$, the Cartan dual of S$^{d+1}$, [\pref{CandH}].\footnote{
This integral occurs in the functional relation for a Selberg \zf, [\pref{Kurokawa}].} A
similar representation holds also for spinors. The Plancherel measure plays a basic role  in
the hyperbolic cylinder approach, [\pref{DandM}].

Note that the product lacks the terms corresponding to critical forms, $\be=\pm t$. At
these points the derivative takes the value $(-1)^p$ except for $\be=0$ (\ie a zero order
operator) when it equals $2(-1)^{d/2}$.

Perhaps the representation (\peq{planch})  is not unexpected.  A stiffer test would be the
corresponding evaluation of functional determinants in odd dimensions. This is not
attempted here.

\section{ \bf 10. The complete field theory $\ze(0)$}
Having the single (coexact) $p$--form quantities, it is possible to assemble a
ghosts--for--ghosts sum to get the complete physical, gauge invariant $p$--form free
energy for the correct number of degrees of freedom, presumably $\comb {d-2}p$.

Assuming a standard Lagrangian formulation, I simply write down the Obukhov
construction in terms of the unconstrained quantities. These could be eliminated in favour
of just the coexact ones, [\pref{CandA}], but I won't make use of this. (See the
Appendix.) The expression is, \eg\ [\pref{CandT2}],[\pref{CandA}],
  $$
  \caF(p,q,d,k)=\sum_{l=0}^p(-1)^l(1+l)\gF (p-l,q,d,k)\,,
  \eql{gfg}
  $$
where $\caF$  and $\gF$ stand, generically, for free energies  that is for a conformal
anomaly, or for a functional determinant, or for an interpolation between these, if such is
possible

This is my construction of the quantised theory. $\caF$ is the gauge invariant, `physical'
free energy constructed out of the `geometric', free--form quantity, $\gF$, which is, in the
case under discussion here, the conformal anomaly, $C^{tot}$.

The manifold could be the $q$--lune with absolute or relative boundary conditions, but I
consider just the $q$--deformed sphere for brevity.

As a preliminary, simple check, I remark that (\peq{gfg})  reproduces the well known
conformal scalar anomalies on the sphere.

The end results are again rational functions of $q$ and $k$ for given $p$ and $d$. I  now
display some $q$--sphere conformal anomalies, and make some observations.

Taking the forms in 6 dimensions on the $q$--sphere, as typical, I find,
  $$\eqalign{
  \caF(2,q,6,k)={1\over5040}&k\big(2q^5-7(2k^2-3)q^3+42(k^2-1)(k^2-4)q\big)\cr
  &-{1\over90}k^3(3k^2+35)+{k\over 5040q}(12k^6-126k^4+336k^2-191)\cr
  \caF(1,q,6,k)={1\over7560}&k\big(2q^5-7(2k^2-3)q^3+42 (k^2-1) (k^2-4 )q\big)\cr
  &-{1\over180}k^3 (3k^2+5 )+{k\over7560q}(12 k^6-126 k^4+336 k^2-191)\cr
  \caF(0,q,6,k)={1\over30240}&\big(k(2q^5-7(2k^2-3)q^3+42(k^2-1)(k^2-4)q\big)\cr
  &+{k\over30240q}(12k^6-126k^4+336k^2-191)\,.
  }
  \eql{6d2}
  $$

The coefficient of $q$ is, generally,
  $$
  {1\over2}{\pa^2\over\pa q ^2} \big(q\,\caF(p,q,d,k)\big)\bigg|_{q=0}
  ={1\over6(d-1)!}\comb{d-2}p\prod_{j=0}^{d/2-1}(k^2-j^2)\,.
  $$
which exhibits the conformal vanishing at allowed $k$ values.

One also sees in this particular formula evidence of the expected  reduction in propagating
components. In fact, calculation produces for the top two components the values,
  $$\eqalign{
  \caF(d,q,d,k)&=(d+2)k\cr
  \caF(d-1,q,d,k)&=(-1)^p d \,k\,.
  }
  $$
One might expect a physical Hodge duality, $p\to d-2-p$, under which these two
components have no dual, and so should vanish, or have a topological or cohomological
value.

More generally, calculation reveals that the Hodge `anomaly' is,
  $$
  \caF(p,q,d,k)-\caF(d-2-p,q,d,k)=(-1)^p\,2k\big(p-{d\over2}+1\big)\,,
  \eql{hanom}
  $$
antisymmetrical about the $k=1$ critical value.

A similar circumstance occurs in the $p$-form calculations in [\pref{dowCTpform}] and
more especially in [\pref{Raj}]. Further analysis of Hodge duality is given in the Appendix
justifying (\peq{hanom}). The paper [\pref{DMW}] contains a more thorough investigation
into duality and (\peq{hanom}) fits in with their results.

It is important to note that when $k=1$, \ie for a usual second order operator, the
expressions (\peq{6d2}) do not agree with those in [\pref{dowCTpform}] and [\pref{Raj}]
even in the critical case. This should be expected since even if the basic $p$--form is
critical, the ghosts are not, having different  operators  in the two cases. I therefore list
some numbers for the conformal anomaly on the full sphere, for the present theory when
$k=1$ and $p=d/2-1$, the critical, conformal value, ($p$ runs from 0 to 7), \footnote{ I
also record some values on the present scheme  for $k=2$, (\ie $p=d/2-2$), $
[14/45,$\break$-326/945,113/315,-171277/467775].$}
  $$\eqalign{
  &\caF_{here}\cr
  &=\bigg[{1\over3},\,-{16\over45},\,{229\over630},\,-{1042\over2835},\,
  {276929\over748440},\,-{45201643\over121621500},\,{108829363\over291891600},\,
  -{121702602491\over325641566250}\bigg]\,.\cr
  }
  \eql{crit1}
  $$

These are to be compared with the corresponding values in [\pref{CandA}] for $p=0$ to
$p=7$, copied \footnote{ I have extended by the $p$=0  and $p=$7 values. The $p=8$
value is  9098897310129059/\break 2771861011920000.} here for convenience,
  $$\eqalign{
 &F_{standard}\cr
 &=\bigg[{1\over3},\, -{31\over45},\,{221\over210},\,-{8051\over5670},
 \,{1339661\over748440},\,
  -{525793111\over243243000},\,{3698905481\over1459458000}
  ,\,-{7576167103513\over2605132530000}\bigg]\,.\cr
  }
  \eql{canda}
  $$

It is remarkable that contact with these earlier results  can be made by the following
observation.

It will be found that an alternating sum of critical values, (\peq{crit1}),
  $$
   F=\sum_{j=0}^p (-1)^j\,\caF(p-j,1,2p+2-2j,1)\,,
   \eql{altsum}
  $$
reproduces the standard ones, (\peq{canda}). Conversely, adding two adjacent values in
(\peq{canda}) yields a term in (\peq{crit1}). I have no deep justification for this
numerical fact. However, it seems reasonable to generalise it by defining,
   $$
   F_{crit}(p,q,k)\bigg|_{d=2p+2k}\equiv\sum_{j=0}^p (-1)^j\,\caF(p-j,q,2p+2k-2j,k)\,,
   \eql{fcrit}
  $$
just for critical $p$--forms of a certain propagation order, $2k$.

In the present calculational scheme, it is not possible to find $F_{crit}$ as a polynomial in
$k$. Case by case evaluation is needed. I list a few values on the full sphere for $p=0$ to
$p=3$, and $k=2$ to $k=4$,
$$\eqalign{
F_{crit}(p,1,2)&={14\over45},\,{124\over189},\,{137\over135},\,{645982\over467775}\cr
\noalign{\vskip 3pt}
F_{crit}(p,1,3)&={41\over140},\,{437\over700},\,{1873\over1925}
,\,{466497\over350350}\cr
\noalign{\vskip3pt}
F_{crit}(p,1,4)&={3956\over14175},\,{56008\over93555},\,{725692\over773955},\,
{493493584\over383107725}\,.\cr
}
   $$

I do not discuss these quantities any further at present. Lengthier comments can be found
in section 12 concerning (\peq{crit1}) and (\peq{altsum}).

For the full sphere, I have not succeeded in finding  a form like (\peq{planch}) for the
gauge invariant conformal anomaly so I simply give a few of the resulting
$k$--polynomials
 $$\eqalign{
\caF(1,1,4,k)&={1\over90}k^3(3k^2-35)\cr
\caF(2,1,4,k)&={1\over180}k(3k^4-5k^2+360)\cr
\caF(1,1,6,k)&={1\over3780}k^3(6k^4-105k^2+161)\cr
\caF(2,1,6,k)&={1\over1260}k^3(3k^4-63k^2+518)\cr
\caF(3,1,8,k)&={1\over45360}k^3(5k^6-198k^4+2709k^2-19188)\,.
}
   $$

\mgn{pforment3.wxm}

\section{\bf 11. Casimir energy on the Einstein cylinder}

In earlier works, [\pref{dowqretspin}], [\pref{dowgjmsren}], it was noticed that the
coefficient of the $1/q$ term in the free energy, $\caF$, (conformal anomaly) was minus
twice the Casimir energy of the field on the Einstein Universe\footnote{ Branson,
[\pref{Branson}], is mostly concerned with propagation on this.} (generalised cylinder),
T$\times$S$^{d-1}$. A direct confirmation of this in the present set up would provide
evidence that the system is consistent. This task will be undertaken at another time. For
now, as a prediction, I list some vacuum energies, $\caE_0(p,d,k)$, obtained on the basis
of the above correspondence,
  $$\eqalign{
\caE_0(p,4,k)&=-{1\over720}\comb2p k(6k^4-20k^2+11)\cr
\caE_0(p,6,k)&=-{1\over60480}\comb4pk(12k^6-126k^4+336k^2-191)\cr
\caE_0(p,8,k)&=-{ 1\over3628899}\comb6p k(10k^8-240k^6+1764k^4-4320k^2+2497)\,.\cr
    }
    \eql{ezero}
  $$
The scalar $p=0$ values agree with those computed in [\pref{dowqretspin}]. Our present
results verify that the general $p$-form energies are simply weighted with the dynamical
degrees of freedom, $\comb{d-2}p$. Some graphs of the $k$--dependence are to be found
in [\pref{dowqretspin}].

The $p>0$ results do not agree with the existing ones, as anticipated. It would be
expected, however, that the {\it critical} values are related by (\peq{fcrit}), at least for
$k=1$, the only case available up to now. To make this evident, I again give some
numbers for the critical forms calculated from, say, (\peq{ezero}) at $k=1$ (displayed for
$p=0$ to 6),
  $$\eqalign{
    &\caE_0^{crit}=\cr
    \noalign{\vskip4truept}
    &\bigg[  -{1\over12},\,{1\over120},\,-{31\over10080},\,{289\over181440},\,
  -{2219\over2280960},\,{6803477\over10378368000},\,
-{3203699\over6793113600}\ldots\bigg]
  }
  \eql{cashere}
  $$
and quote an extended list of the corresponding ones, $E_0$, given in [\pref{GKT}],
computed directly on the Einstein cylinder with standard `Maxwell' theory. An efficient
general formula can be found in [\pref{dowqretspin}]. For $p=1$ to 7, one has,
   $$\eqalign{
   &E_0^{crit}=\cr
   \noalign{\vskip4truept}
   &\bigg[{11\over120},\,-{191\over2016},\,{2497\over25920}
   ,\,-{14797\over152064},\,{92427157\over943488000}
   ,\,-{36740617\over373248000},\,{61430943169\over621831168000},\,\ldots\bigg]\,
   }
   \eql{casstand}
   $$

As before, an alternating sum of (\peq{cashere}) yields (\peq{casstand}) and the addition
of two adjacent terms in (\peq{casstand}) gives one in (\peq{cashere}). Using this, the
numbers in (\peq{casstand}) can be extended correctly to the left by $-1/12$ for $p=0$
and $0$ for $p=-1$ while (\peq{cashere}) can be extended to the right by
73691749/207277056000. See Appendix B for a more analytical connection between
(\peq{cashere}) and (\peq{casstand}).

\begin{ignore}

  \section{\bf Can the degeneracies be simplified.??????}

$$
  d_a(p,\si,q)={(-1)^{p+1}\over \si^{p+1}(1-\si)^{d-1}(1-\si^q)}\sum_{r=p+1}^d
  (-1)^r\,e_r(\si^q,\si,\ldots,\si)\,,
  \eql{deea1}
  $$
Explicitly
  $$
  e_r(\si^q,\si,\ldots,\si)=\comb{d-1}r\,\si^r+\comb{d-1}{r-1}\,\si^{r-1+q}\,,
  \eql{esf}
  $$

Try first $q=1$ (hemisphere) and show equivalence with [\pref{DandKi}] which is (P-forms
II section 8.) and start from this side,
  $$\eqalign{
  g_a(p,\si)&=\sum_{m=p+1}^d\comb{m-1}{p}{1\over(1-\si)^m}\cr
  &={1\over(1-\si)^d}\sum_{m=p+1}^d\comb{m-1}{p}(1-\si)^{d-m}\cr
  &={1\over(1-\si)^d}\sum_{m=p+1}^d\comb{m-1}{p}\sum_{s=0}^{d-m}
  \comb{d-m}{s}(-1)^s\si^s\cr
   &={1\over(1-\si)^d}\sum_{m=1}^d\comb{m-1}{p}\sum_{s=0}^{\infty}
  \comb{d-m}{s}(-1)^s\si^s\cr
  &={1\over(1-\si)^d}\sum_{s=0}^\infty\sum_{m=1}^{d}\comb{m-1}{p}
  \comb{d-m}{s}(-1)^s\si^s\cr
  }
  $$

Vandermonde is
  $$
\sum_{\mu=0}^n\comb{\mu}{j}\,\comb{n-\mu}{k-j}=\comb{n+1}{k+1}
  $$
 Set $\mu=m-1$, $j=p$, $k-j=s$, $n=d-1$ \ie $k=s+p$

 Hence
   $$\eqalign{
  g_a(p,\si)
  &={1\over(1-\si)^d}\sum_{s=0}^\infty\sum_{m=1}^{d}\comb{m-1}{p}
  \comb{d-m}{s}(-1)^s\si^s\cr
   &={1\over(1-\si)^d}\sum_{s=0}^\infty\comb{d}{s+p+1}(-1)^s\si^s\cr
  }
  \eql{geea}
  $$
Now, at $q=1$,
 $$\eqalign{
 d_a(p,\si.1)={(-1)^{p+1}\over\si^{p+1}(1-\si)^d}\sum_{r=p+1}^d(-1)^r\comb dr\si^r
}
 $$
Now set $r=s+p+1$ and the two quantities are identical, the upper $r$ limit being
governed by the vanishing of the binomial.

Look at each term in (\peq{esf}) separately. First term bit like above but with $d\to d-1$
($r=d$ term is zero)

$$\eqalign{
 d_a(p,\si,1)\big |_{d\to d-1}&={(-1)^{p+1}\over\si^{p+1}
 (1-\si)^{d-1}}\sum_{r=p+1}^{d-1}(-1)^r\comb {d-1}r\si^r\cr
 &=\sum_{m=p+1}^{d-1}\comb{m-1}{p}{1\over(1-\si)^m}\cr
}
\eql{firstt}
 $$
using the previous identity.

Hence first term is, putting in the overall factors
  $$
{1\over(1-\si^q)}\sum_{m=p+1}^{d-1}\comb{m-1}{p}{1\over(1-\si)^m}
\eql{firstt2}
  $$

Try to reverse the procedure leading to (\peq{geea}). The first term has been dealt with in
(\peq{firstt2}).

Second term
  $$\eqalign{
  \sum_{r=p+1}^d(-1)^r&\comb{d-1}{r-1}\,\si^{r-1+q}=
 -\si^q \sum_{\rho=p}^{d-1}(-1)^\rho\comb{d-1}{\rho}\,\si^{\rho}\cr
 &=-(-1)^p\si^q\sum_{s=0}^{d-1-p}(-1)^s\comb{d-1}{s+p}\,\si^{s+p}\cr
 &=-(-1)^p\si^{q+p}\sum_{s=0}^{\infty}(-1)^s\comb{d-1}{s+p}\,\si^{s}\cr
&=-(-1)^p\si^{q+p}\sum_{s=0}^{\infty}\sum_{m=1}^{d-1}(-1)^s
\comb{m-1}{p-1}\comb{d-1-m}{s}\,\si^{s}\cr
&=-(-1)^p\si^{q+p}\sum_{m=1}^{d-1}
\comb{m-1}{p-1}\sum_{s=0}^{\infty}\comb{d-1-m}{s}\,(-1)^s\si^{s}\cr
&=-(-1)^p\si^{q+p}\sum_{m=p}^{d-1}
\comb{m-1}{p-1}\sum_{s=0}^{d-1-m}\comb{d-1-m}{s}\,(-1)^s\si^{s}\cr
&=-(-1)^p\si^{q+p}\sum_{m=p}^{d-1}
\comb{m-1}{p-1}(1-\si)^{d-1-m}\equiv \si^{p+q} X\cr
  }
  $$

Notes:

Putting $s=\rho-p$. The finite upper limit occurs when $s+p=d-1$.

Vandermonde is
  $$
\sum_{\mu=0}^n\comb{\mu}{j}\,\comb{n-\mu}{k-j}=\comb{n+1}{k+1}
  $$
 \vglue 20truept
$k+1=s+p$, $\mu=m-1$, $n-\mu=d-1-m$,  $j=p-1$, $n+1=d-1$, $k-j=s$. Therefore
$n=d-2$ and $k=s+p-1$. Checks.

Outside factors

$$
{(-1)^{p+1}\over \si^{p+1}(1-\si)^{d-1}(1-\si^q)}
$$
Therefore $\si^p$ cancels to leave $1/\si$.  Leave this outside. Signs combine to $+1$.
The factor $(1-\si)^{d-1}$ cancels

Make same split as before to cancel $(1-\si^q)$ on bottom. Therefore write
$\si^q\si^pX=-(1-\si^q)\si^pX+\si^pX$. Including the outside factors get
  $$\eqalign{
  {1\over(1-\si^q)}\sum_{m=p+1}^{d-1}\comb{m-1}{p}
  {1\over(1-\si)^m}+&{1\over\si(1-\si^q)}\sum_{m=p}^{d-1}
\comb{m-1}{p-1}{1\over(1-\si)^{m}}\cr
&-{1\over\si}\sum_{m=p}^{d-1}
\comb{m-1}{p-1}{1\over(1-\si)^{m}}
}
  $$

Multiplying by $\si^{a+1}$, Mellin gives ($a=(d-1)/2$,
  $$\eqalign{
  \sum_{m=p+1}^{d-1}\comb{m-1}{p}
  \ze_\caB(s,a+1\mid q,{\bf1}_m)+&\sum_{m=p}^{d-1}
\comb{m-1}{p-1}\ze_\caB(s,a\mid q,{\bf1}_m)\cr
&-\sum_{m=p}^{d-1}
\comb{m-1}{p-1}\ze_\caB(s,a\mid{\bf1}_m)
}
  $$

Check for $q=1$. All parameter are 1. Recursion is
  $$
  \ze_\caB(s,a+1\mid{\bf1}_{m+1})=\ze_\caB(s,a\mid{\bf1}_{m+1})
  -\ze_\caB(s,a\mid{\bf1}_{m})
  $$

Answer should be,
  $$
\sum_{m=p+1}^d\comb{m-1}p\ze_B(s,a+1\mid{\bf1}_m)
$$

\section{\bf8. $q\to0$ limit and Casimir energies}

Alternative expressions for the free energies to those derived directly from the \zf\
(\peq{zet4}) are more advantageous in order to make contact with the results in
[\pref{dowqretspin}]. These are obtained by re-organising the degeneracies in the cylinder
kernel, (\peq{ck})\footnote{ In other language this is a re-organisation of the
$q$--series.} Briefly, the binomial coefficients in the symmetric function, (\peq{esf}),
needed for the degeneracy, (\peq{deea}), are split into two factors using the
Vandermonde formula. This enables the sum to be taken over the {\it dimension} of the
Barnes \zfs.

I do not give the algebraic details but just present the alternative expression to
(\peq{zet4}) for the simple \zf,
  $$\eqalign{
  \ze_a(s,a,p,q)=\sum_{m=p+1}^{d-1}\comb{m-1}{p}
  \ze_\caB(s,a+1\mid q,{\bf1}_m)+&\sum_{m=p}^{d-1}
\comb{m-1}{p-1}\ze_\caB(s,a\mid q,{\bf1}_m)\cr
&-\sum_{m=p}^{d-1}
\comb{m-1}{p-1}\ze_\caB(s,a\mid{\bf1}_m)
}
  $$

This can be used for formal as well as calculational purposes since it displays the
$p$--dependence more simply. Further, because of the vanishing of the binomial
coefficients, the lower summation limits can be taken to be $0$ or $-1$ \etc, at will. First
$s$ to 0 to obtain,
  $$\eqalign{
  q\,\ze_a&(0,a,p,q)=-\sum_{m=p+1}^{d-1}{(-1)^m\over(m+1)!}\comb{m-1}{p}
  B^{(m+1)}_{m+1}(a+1\mid q,{\bf1}_m) \cr-
&\sum_{m=p}^{d-1}{(-1)^m\over(m+1)!}
\comb{m-1}{p-1}B^{(m+1)}_{m+1}(a\mid q,{\bf1}_m)\cr
&-q\,\sum_{m=p}^{d-1}{(-1)^m\over m!}
\comb{m-1}{p-1}B^{(m)}_{m}(a\mid {\bf1}_m)
}
  $$
Therefore at $q=0$,
  $$\eqalign{
  q\,\ze_a&(0,a,p,q)\bigg|_{q=0}=-\sum_{m=p+1}^{d-1}{(-1)^m\over(m+1)!}\comb{m-1}{p}
  B^{(m)}_{m+1}(a+1\mid {\bf1}_m) \cr-
&\sum_{m=p}^{d-1}{(-1)^m\over(m+1)!}
\comb{m-1}{p-1}B^{(m)}_{m+1}(a\mid {\bf1}_m)\,.\cr
}
\eql{z0}
  $$

To get the free energy, the average (\peq{pca}) has to be taken.

At this point I give the expression for the Maxwell (conformal) Casimir energy on
R$\times$S$^{d-1}$, $p=d/2-1$
 $$\eqalign{
  E^M_0(p)
  &=\sum_{m=p+1}^{2p+1}{(-1)^m\over(m+1)!}\comb{m-1}p
  \,B^{(m)}_{m+1}\big(p+1\big)\,,\cr
  }
  \eql{maxen}
  $$

The arguments of the $b$s in (\peq{z0}) are $(d-1)/2+1+1/2=d/2+1$,
$(d-1)/2+1-1/2=d/2$,  and $d/2$, $d/2-1$ for the Maxwell case.

\end{ignore}
\section{\bf12.  Comments and Conclusion}
One aspect (the conformal anomaly) of a quantised field propagating according to the
Branson--Gover operators on the sphere has been investigated. Even for second order
propagation the results differ from the conventional ones for $p>0$ because of the
differring ghost system. Both the present theory and the conventional one display the
correct degrees of freedom, $\comb {d-2}p$.

Numerically it is observed that the two sets of values are simply related when the forms
are `critical' \ie when the derivative order, form order and dimension satisfy $2k=d-2p$.
This relation, (\peq{altsum}), has the appearance of the sum involving edge mode
`corrections' that yields, for example, the Maxwell entanglement entropy as the conformal
anomaly [\pref{DMW}]. \footnote{ A similar interpretation of the relation between the
Casimir energies is not so obvious. More information is given in Appendix B.} Why this
should be is not entirely clear.

The numbers (\peq{crit1}) agree with those in Table 2 of [\pref{DandM}]. They show that
both theories deliver the entanglement entropy devoid of any corrections.

The numbers for any $p$ in [\pref{DandM}] were constructed using just the coexact forms,
as an extrapolation from the detailed gauge calculations for $p=2$.

Also in connection with the hyperbolic approach in [\pref{DandM}], it seems that the
Branson operators, for $k=1$, saturate the Breitenlohner--Freedman bound.

A clean mathematical result is that the unconstrained $p$--form conformal anomaly on
S$^d$ can be expressed as an integral over the (continued) Plancherel measure for
coexact $p$--forms on the odd hyperbolic space, H$^{d+1}$, indicating connections with
a form Selberg \zf, [\pref{Kurokawa}].

It is also shown that the entanglement entropy for the sphere equals minus the conformal
anomaly, technically because the conformal anomaly on the $q$--deformed sphere is an
extremum at the round value, $q=1$, (but only for allowed values of $k$).

For brevity, I have not extended the analysis beyond that involving the first derivatives
with respect to $q$.

Further investigations include  finding the functional determinants in odd dimensions
where one would want to discuss bounded manifolds like the hemisphere in more detail.

\section{\bf Appendix A. Duality}

There is some general interest in the validity of Hodge duality. In the present situation a
simple analysis can be given starting from the ghost sum, (\peq{gfg}), as in most other
treatments, including the historic ones.  For variety, I use the alternative formulation in
terms of the coexact conformal anomalies on the $q$--lune, [\pref{CandA}],

$$
  \caF^b(p)=\sum_{l=0}^p(-1)^{p+l}\gF^b_{CE}(l)+(-1)^p\,(p+1)k\de^{br}
  \eql{gfg2}
  $$
and have dropped the inessential arguments, $d,q,k$.

The final term is a zero mode effect which exists only for relative conditions, $b=r$.

One makes the duality replacement $p\to d-p-2$ in this expression and, to begin with, I
add the absolute and relative values so that the system is a $p$-form on the periodic
lune. To make the notation more streamlined for this discussion, I denote the single lune
by $\man$ and its periodic double by $2\man$. In terms of \zfs, the relation is,
 $$\eqalign{
  \ze(2\man,p)&=\ze^a(\man,p)+\ze^r(\man,p)\cr
  &=\ze^a(\man,p)+\ze^a(\man,d-1-p)\,.
  }
 $$

The ghost sum then reads,
  $$\eqalign{
  \caF(2\man,p)=&\sum_{l=0}^p(-1)^{p+l}\gF_{CE}(2\man,l)+(-1)^p\,(p+1)k\cr
  &=(-1)^{d-1}\sum_{l=d-1}^{d-1-p}(-1)^{p+l}\gF_{CE}(2\man,l)+(-1)^p\,(p+1)k\,,\cr
  }
  \eql{gfg3}
  $$
on redefining $l$ and using coexact $a\leftrightarrow r$ duality \ie
$\gF_{CE}(2\man,l)=\gF_{CE}(2\man,d-1-l)$. This is an identity.

Now set $p\to d-2-p$ in the second term to get
 $$\eqalign{
  \caF(2\man,d-2-p)&=-\sum_{l=p+1}^{d-1}(-1)^{p+l}\gF_{CE}(2\man,l)
  +(-1)^p\,(d-1-p)k\cr
  }
 $$
so that the difference in Hodge duals is,
  $$\eqalign{
  \caF(2\man,p)-\caF(2\man,d-2-p)&=\sum_{l=0}^{d-1}(-1)^{p+l}\gF_{CE}(2\man,l)
  +(-1)^p\,(2p+2-d)k\cr
  &=(-1)^p2(p+1-d/2)k
  }
 $$
which is the value in the main text.

For the more general single $q$-lune with boundary conditions, a similar manipulation
reveals the Hodge dual relation,\mgn{list3. Check sign}
  $$\eqalign{
  \caF^a(\man,p)-\caF^r(\man,d-2-p)
  &=-(-1)^p(3d/2-1-p)
  }
 $$

 Also\mgn{list2},
   $$
  \caF^a(\man)-\caF^r(\man)=-(-1)^p2(p+1)+\de_{pd}
  $$

Therefore \mgn{list7} one arrives at the curiously asymmetric Hodge--like dualities,
  $$\eqalign{
  \caF^r(\man,p)-\caF^r(\man,d-2-p)
  &=(-1)^p(2p+2-(3d/2-1-p))k\cr
  &=3(-1)^p(p+1-d/2)k-k \de_{pd}\cr
  }
 $$
and
$$\eqalign{
  \caF^a(\man,p)-\caF^a(\man,d-2-p)
  &=-(-1)^p(p+1-d/2)k+k\de_{pd}\cr
  }
 $$

  $$\eqalign{
  \caF^r(\man,d-2-p)-\caF^r(\man,p)
  &=3(-1)^{d-2-p}(d-2-p+1-d/2)k\cr
  &=3(-1)^p(d/2-1-p)k\,.
  }
 $$

These relations give the particular values when $p=d-1$ and $p=d$.
\section{\bf Appendix B. More numerology on the Einstein cylinder}

The relation between the Casimir energies (\peq{cashere}) and (\peq{casstand}) can be
made a little more systematic in the following way.

In [\pref{dowqretspin}], the Casimir energy on the Einstein cylinder was obtained
conventionally from the coexact Maxwell \zf\ on the (odd) sphere S$^{d-1}$ that is, the
\zf\ for a critical (conformal) $p$--form ($d=2p+2$). This was obtained from the
generating function found in [\pref{DandKii}] manipulated to produce a sum of Barnes
\zfs\ different in form (but equivalent) to that used here for the auxiliary \zf. The result is
not needed here.

Also in [\pref{DandKii}], using recursions, the $p$--form coexact \zf\ on a  sphere
S$^{d-1}$, was expressed as an alternating sum of conformal  \footnote{ Remember that
the system is conformally covariant in $d$ dimensions not $d-1$.} scalar, $0$--form \zfs,
$\ze^{conf}_0(s)$,  on spheres of different dimensions. In the critical, conformal case, \ie
$p=d/2-1$, it reads,
  $$
       \ze^{CE}(s,p)\bigg|_{S^{2p+1}}=\sum_{j=0}^p(-1)^{j+p} \comb{2j}{j}
       \ze^{conf}_0(s)\bigg|_{S^{2j+1}}\,.
       \eql{dandkir}
  $$

Before proceeding, it should be noted that, on the Einstein cylinder, the single coexact \zf\
is sufficient as the total ghost--for--ghost sum collapses through t\'elescopage,
[\pref{dowzero}].\footnote{ Something similar seems to occur, but more complicatedly in
detail, in the hyperbolic approach, [\pref{DandM}].}

Evaluation of (\peq{dandkir}) at $s=-1/2$ gives a formula for the standard $p$--form
Casimir energy on the Einstein cylinder in terms of conformal scalar Casimir energies on
Einstein cylinders of varying dimensions,
  $$
  E_0^{crit}(p)\big|_{S^{2p+1}}=\sum_{j=0}^p(-1)^{p+j} \comb{2j}{j}
       E_0^{conf}(0)\big|_{S^{2j+1}}\,.
       \eql{eezcrit}
  $$

The conformal scalar Casimir energy on the odd--sphere cylinder was calculated some time
ago, [\pref{ChandD}], with the result,
     $$
     E_0^{conf}(0)\big|_{S^n}=-{1\over(n+1)!}B^{(n)}_{n+1}\big((n-1)/2\big)\,,
     \quad n\,\,{\rm odd}\,.
     $$
The relation between (\peq{cashere}) and (\peq{casstand}) is then expressed by,
  $$
     E^{crit}_0(p)\big|_{S^{2p+1}}=\sum_{j=0}^p (-1)^{p+j}\caE^{crit}_0(j)\,,
     \eql{ecrit}
  $$
where,
  $$
    \caE^{crit}_0(j)\equiv-{1\over 2(j+1)(2j+1)\,j!^2}B^{(2j+1)}_{2j+2}(j)\,,
  $$
which evaluates to the list (\peq{cashere}).

This makes the relation with (\peq{casstand}) more precise but sheds no physical light on
why (\peq{cashere}) should emerge from the system defined on the even $q$--sphere as
$q\to0$. In fact, equation (\peq{ecrit}) could be looked upon as just another way of
computing (\peq{casstand}). Different ways of expressing the $p$--form \zf\ correspond to
different arrangements of the generating function (or `$q$--series' or `character'), no one
way seemingly being more significant than any other.\footnote{ Similar considerations
would hold in the hyperbolic calculation, [\pref{DandM}], regarding manipulations with the
Plancherel measure \eg\ expansion.}

I note that $\caE^{crit}_0(j)$ has the correct number of degrees of freedom for a critical
$j$--form, being equivalent to $\comb {2j}j$ single scalars and that the construction of
(\peq{ecrit}) is an alternating sum over these scalars, which have conformal dimension,
$j$.


 \vskip15truept
 \noin{\bf References.} \vskip5truept
\begin{putreferences}
  \ref{FandS}{Fischmann,M. and Somberg,P., {\it The boundary value problem
  for Laplacian on differential forms and conformally Einstein
  infiniity},{\it J.Gen.Lie Th.Appl.} {\bf 11} (2017) 256,
   ArXiv: 1508. 01511 [math.DG].}
   \ref{DandM}{David,J.R. and Mukherjee,J. {\it Hyperbolic cylinders and Entanglement entropy:
   gravitons, higher spins and $p$--forms.}, ArXiv:2005.08402.}
   \ref{BandC}{Benedetti,V. and Casini,H. {\it Entanglement entropy of linearized
    gravitons in a sphere,} ArXiv:1908.01800.}
  \ref{DandKii}{Dowker,J.S. and Kirsten, K. {\it Spinors and forms on the ball and the
   generalised cone},  {\it Comm. in Anal. and Geom. }{\bf7} (1999) 641,
   ArXiv:hep--th/9608189.}
  \ref{Kurokawa}{Kurokawa,S, {\it Gamma Factors and Plancherel Measures,}
  {\it Proc.Jap. Acad.} {\bf 68} (1992) {256}.}
  \ref{dowCTpform}{Dowker,J.S, {\it R\'enyi entropy and $C_T$ for p-forms on
  even spheres}, ArXiv: 1706. 04574.}
  \ref{Branson}{Branson,T. {\it Group Representations Arising from Lorentz
  Conformal Geometry}, {\it J.Func.Anal.} {\bf 74} (1987) 199. }
  \ref{BandG}{Branson,T. and Gover,A.R. {\it Conformally invariant operators
, differential forms, cohomology and a generalization of $Q$--curvature},
 {\it Comm.Partial
Diff.Equns} {\bf30} (2005)
  1669.}
  \ref{BandT}{Beccaria,M. and Tseytlin,A.A. {\it $C_T$ for higher derivative conformal
  fields and anomalies of (1,0) superconformal 6d theories}, ArXiv:1705.00305.}
  \ref{NandZ}{Nian,J. and Zhou,Y. {\it R\'enyi entropy of free (2,0) tensor multiplet
  and its supersymmetric counterpart}, \prD{93}{2016}{125010}, ArXiv:1511.00313}
  \ref{Huang}{Huang,K--W., {\it Central Charge and Entangled Gauge Fields},
  \prD {92}{2015}{025010}, ArXiv:1412.2730.}
  \ref{NFM}{De Nardo,L., Fursaev,D.V. and Miele,G. {\it}\cqg{14}{1997}{1059}, ArXiv:
  hep-th/9610011.}
  \ref{Fursaev}{Fursaev,D.V.,{\it Entanglement R\'enyi Entropies in Conformal
  Field Theories and Holography},{\it JHEP} 0609:018,2006. ArXiv:1201.1702.}
  \ref{Raj}{Raj,H. {\it A note on sphere free energy of $p$--form gauge theory and
  Hodge duality}, \cqg{34}{2017}{247001}, ArXiv:1611.02507.}
  \ref{DMW}{Donnelly.W, Michel,B. and Wall,A.C. {\it Electromagnetic
  duality and entanglement entropy}, ArXiv:1611.0592.}
  \ref{GKT}{Giombi,S. Klebanov,I.R. and Tan, Z.M. {\it The ABC of Higher--Spin AdS/CFT}.
  {\it Universe} {\bf 4} (2018) 1, ArXiv:1608.07611.}
  \ref{dowzero}{Dowker,J.S. {\it Zero modes, entropy bounds and partition functions},
  \cqg{20}{2003}{L105}, ArXiv:hep-th/0203026.}
  \ref{dowqretspin}{Dowker,J.S. {\it Revivals and Casimir energy for a free Maxwell field
  (spin-1 singleton) on $R\times S^d$ for odd $d$}, ArXiv:1605.01633.}
  \ref{Wunsch}{W\"unsch,V. {\it On Conformally Invariant Differential Operators}, {\it Math.
  Nachr.} {\bf 129} (1989) 269.}
  \ref{BEMPSS}{Buchel,A.,Escobedo,J.,Myers,R.C.,Paulos,M.F.,Sinha,A. and Smolkin,M.
  {\it Holographic GB gravity in arbitrary dimensions}, {\it JHEP} 1003:111,2010,
  ArXiv: 0911.4257.}
  \ref{dowpform1}{Dowker,J.S. {\it $p$--forms on$d$--spherical tessellations}, {\it J.
  Geom. and Phys.} ({\bf 57}) (2007) 1505, ArXiv:math/0601334.}
  \ref{dowpform2}{Dowker,J.S. {\it $p$--form spectra and Casimir energies on spherical
  tessellations}, \cqg{23}{2006}{1}, ArXiv:hep-th/0510248.}
  \ref{dowgjmsren}{Dowker,J.S. {\it R\'enyi entropy and $C_T$ for higher derivative
  scalars and spinors on even spheres}, ArXiv:1706.01369.}
  \ref{GPW}{Guerrieri, A.L., Petkou, A. C. and Wen, C. {\it The free $\si$CFTs},
  ArXiv:1604.07310.}
  \ref{GGPW}{Gliozzi,F., Guerrieri, A.L., Petkou, A.C. and Wen,C.
   {\it The analytic structure of conformal blocks and the
   generalized Wilson--Fisher fixed points}, {\it JHEP }1704 (2017) 056, ArXiv:1702.03938.}
  \ref{YandZ}{Yankielowicz, S. and Zhou,Y. {\it Supersymmetric R\'enyi Entropy and
  Anomalies in Six--Dimensional (1,0) Superconformal Theories}, ArXiv:1702.03518.}
  \ref{OandS}{Osborn.H. and Stergiou,A. {\it $C_T$ for Non--unitary CFTs in higher
  dimensions}, {\it JHEP} {\bf06} (2016) 079, ArXiv:1603.07307.}
  \ref{Perlmutter}{Perlmutter,E. {\it A universal feature of CFT R\'enyi entropy}
  {\it JHEP} {\bf03} (2014) 117. ArXiv:1308.1083.}
   \ref{Norlund}{N\"orlund,N.E. {\it M\'emoire sur les polynomes de Bernoulli}, \am{43}
   {1922}{121}.}
   \ref{dowqretspin}{Dowker,J.S. {\it Revivals and Casimir energy for a free Maxwell field
  (spin-1 singleton) on $R\times S^d$ for odd $d$}, ArXiv:1605.01633.}
   \ref{Dowpiston}{Dowker,J.S. {\it Spherical Casimir pistons}, \cqg{28}{2011}{155018},
   ArXiv:1102.1946.}
  \ref{Dowchem}{Dowker,J.S. {\it Charged R\'enyi entropy for free scalar fields}, \jpa{50}
  {2017}{165401}, ArXiv:1512.01135.}
  \ref{Dowconfspins}{Dowker,J.S. {\it Effective action of conformal spins on spheres
  with multiplicative and conformal anomalies}, \jpa{48}{2015}{225402}, ArXiv:1501.04881.}
  \ref{Dowhyp}{Dowker,J.S. {\it Hyperspherical entanglement entropy},
  \jpa{43}{2010}{445402}, ArXiv:1007.3865.}
  \ref{dowrenexp}{Dowker,J.S.{\it Expansion of R\'enyi entropy for free scalar fields},
   ArXiv:1412.0549.}
     \ref{CaandH}{Casini,H. and Huerta,M. {\it Entanglement entropy for the $n$-sphere},
     \plb{694}{2010}{167}.}
   \ref{Apps}{Apps,J.S. {\it The effective action on a curved space and its conformal
     properties} PhD thesis (University of Manchester, 1996).}
   \ref{Dowcen}{Dowker,J.S., {\it Central differences, Euler numbers and symbolic methods},
 \break ArXiv:1305.0500.}
 \ref{KPSS}{Klebanov,I.R., Pufu,S.S., Sachdev,S. and Safdi,B.R.
    {\it JHEP} 1204 (2012) 074.}
 \ref{moller}{M{\o}ller,N.M. \ma {343}{2009}{35}.}
 \ref{BandO}{Branson,T., and  Oersted,B \jgp {56}{2006}{2261}.}
  \ref{BaandS}{B\"ar,C. and Schopka,S. {\it The Dirac determinant of spherical
     space forms},\break {\it Geom.Anal. and Nonlinear PDEs} (Springer, Berlin, 2003).}
 \ref{EMOT2}{Erdelyi, A., Magnus, W., Oberhettinger, F. and Tricomi, F.G. {
  \it Higher Transcendental Functions} Vol.2 (McGraw-Hill, N.Y. 1953).}
 \ref{Graham}{Graham,C.R. SIGMA {\bf 3} (2007) 121.}
  \ref{Morpurgo}{Morpurgo,C. \dmj{114}{2002}{477}.}
      \ref{DandP2}{Dowker,J.S. and Pettengill,D.F. \jpa{7}{1974}{1527}}
 \ref{Diaz}{Diaz,D.E. {\it Polyakov formulas for GJMS operators from AdS/CFT},
 {\it JHEP} {\bf 0807} (2008) 103.}
    \ref{DandD}{Diaz,D.E. and Dorn,H. {\it Partition functions and double trace
    deformations in AdS/CFT}, {\it JHEP} {\bf 0705} (2007) 46.}
    \ref{AaandD}{Aros,R. and Diaz,D.E. {\it Determinant and Weyl anomaly of
     Dirac operator: a holographic derivation}, ArXiv:1111.1463.}
  \ref{CandA}{Cappelli,A. and D'Appollonio, G. {\it On the trace anomaly as a measure
  of degrees of freedom}, \pl{487B}{2000}{87}.}
  \ref{CandT2}{Copeland,E. and Toms,D.J. {\it Quantized antisymmetric tensor
  fields and self-consistent dimensional reduction in higher--dimensional
  spacetimes}, \cqg {3}{1986}{431}.}
   \ref{Allais}{Allais, A. {\it JHEP} {\bf 1011} (2010) 040.}
     \ref{Tseytlin}{Tseytlin,A.A. {\it On Partition function and Weyl anomaly of
     conformal higher spin fields} ArXiv:1309.0785.}
     \ref{KPS2}{Klebanov,I.R., Pufu,S.S. and Safdi,B.R. {\it JHEP} {\bf 1110} (2011) 038.}
    \ref{CaandWe}{Candelas,P. and Weinberg,S. \np{237}{1984}{397}.}
     \ref{ChandD}{Chang,P. and Dowker,J.S. {\it Vacuum energy on orbifold
     factors of spheres}, \np{395}{1993}{407}, ArXiv:hep-th/9210013.}
 \ref{Steffensen}{Steffensen,J.F. {\it Interpolation}, (Williams and Wilkins,
    Baltimore, 1927).}
     \ref{Barnesa}{Barnes,E.W. {\it Trans. Camb. Phil. Soc.} {\bf 19} (1903) 374.}
    \ref{DowGJMS}{Dowker,J.S. {\it Determinants and conformal anomalies of
    GJMS operators on spheres}, \jpa{44}{2011}{115402}.}
    \ref{Dowren}{Dowker,J.S. {\it R\'enyi entropy on spheres}, \jpamt {46}{2013}{2254}.}
 \ref{MandD}{Mansour,T. and Dowker,J.S. {\it Evaluation of spherical GJMS determinants},
 2014, Submitted for publication.}
 \ref{GandK}{Gubser,S.S and Klebanov,I.R. \np{656}{2003}{23}.}
     \ref{Dow30}{Dowker,J.S. \prD{28}{1983}{3013}.}
     \ref{Dowcmp}{Dowker,J.S. {\it Effective action on spherical domains},
      \cmp{162}{1994}{633}, ArXiv:hep-th/9306154.}
     \ref{DowGJMSE}{Dowker,J.S. {\it Numerical evaluation of spherical GJMS operators
     for even dimensions} ArXiv:1310.0759.}
       \ref{Tseytlin2}{Tseytlin,A.A. \np{877}{2013}{632}.}
   \ref{Tseytlin}{Tseytlin,A.A. \np{877}{2013}{598}.}
  \ref{Dowma}{Dowker,J.S. {\it Calculation of the multiplicative anomaly} ArXiv: 1412.0549.}
  \ref{CandH}{Camporesi,R. and Higuchi,A. {\it The Plancherel measure for $p$--forms
  in real Hyperbolic space}. {\it J.Geom. and Physics}
  {\bf 15} (1994) 57.}
  \ref{Allen}{Allen,B. \np{226}{1983}{228}.}
  \ref{Dowdgjms}{Dowker,J.S. \jpamt{48}{2015}{125401}.}
  \ref{Dowsphgjms}{Dowker,J.S. {\it Numerical evaluation of spherical GJMS determinants
  for even dimensions}, ArXiv:1310.0759.}
\end{putreferences}
\bye